\documentclass[aps,prb,twocolumn,showpacs,preprintnumbers,amsmath,amssymb]{revtex4}

\usepackage[dvips,final]{graphicx}
\usepackage{dcolumn}
\usepackage{bm}
\usepackage[usenames]{color}

\usepackage{amsmath}	

\newcommand{\bra}[1]{\left<#1\right|}
\newcommand{\ket}[1]{\left|#1\right>}

\begin{document}

\title{Photoinduced magnetic bound state in itinerant correlated electron system \\
with spin-state degree of freedom}
\author{Yu~Kanamori$^{1}$, Jun~Ohara$^{1, 2, \dagger}$, and Sumio~Ishihara$^{1, 2}$} 
\address{$^1$Department of Physics, Tohoku University, Sendai 980-8578, Japan}
\address{$^2$Core Research for Evolutional Science and Technology (CREST), Sendai 980-8578, Japan}

\date{\today}
\begin{abstract}
Photo-excited state in correlated electron system with spin-state degree of freedom is studied. 
We start from the two-orbital extended Hubbard model where energy difference between the two orbitals is introduced. 
Photo-excited metastable state is examined based on the effective model Hamiltonian derived by the two-orbital Hubbard model. 
Spin-state change is induced by photo-irradiation in the low-spin band insulator near the phase boundary. 
High-spin state is stabilized by creating a ferromagnetic bound state with photo-doped hole carriers. 
An optical absorption occurs between the bonding and antibonding orbitals inside of the bound state. 
Time-evolution for photo-excited states is simulated in the time-dependent mean-field scheme. 
Pair-annihilations of the photo-doped electron and hole generate the high-spin state in a low-spin band insulator. 
We propose that this process is directly observed by the time-resolved photoemission experiments. 
\end{abstract}

\maketitle

\section{introduction}

Optical properties and photo-induced phenomena in solids are one of the attractive themes in recent solid state physics. 
In particular, correlated electron system is one of the main targets for photo-induced exotic phenomena. 
Because of strong electron-electron interaction and multi-degrees of freedom, e.g. spin, charge, orbital and so on, 
a number of electronic and structural phases are realized under a subtle balance of interactions.~\cite{Maekawa}
By irradiation of intensive laser pulse into one of the phases, 
a system is transferred into different phase transiently or permanently. 
This is termed photo-induced phase transition (PIPT) phenomena.~\cite{Nasu}
Nowadays, a number of experimental and theoretical studies have been done in PIPT phenomena in transition-metal oxides,~\cite{Fiebig,Cavalleri,Okamoto,Matsueda,Kanamori1}
low-dimensional organic salts,~\cite{Iwai, Chollet, Tajima, Yonemitsu} and others. 

Among the multi-degrees of freedom, spin-state degree of freedom has attracted much attention from view point of optical manipulation of magnetism. In a certain magnetic ion, different magnitude of the spin angular momentum is realized by changing external fields, such as temperature, pressure and photons. This is termed the spin-state transition and is caused by a competition between the crystalline field splitting and the Hund's coupling. 
A well known example of the photo-induced spin state change is seen in the so-called spin cross-over complexes, such as Prussian blue analogue complex.~\cite{Sato1, Bleuzen, Escax, Sato2}
Here, photons bring about a charge transfer from the neighboring Fe ions to Co ions, associated with the spin-state change in Co ions from the low-spin (LS) state to the high-spin (HS) one. 
A main mechanism of the cooperative spin-state transition in a series of materials is supposed to be the elastic interaction;~\cite{Willenbacher,Tchougreeff,Nishino,Miyashita}
a local volume change of a metal-ligand cluster propagates over a crystal lattice. 

Another material where photo-induced spin-state change is realized is the perovskite cobaltites 
$R_{1-x}$$A_x$CoO$_3$ ($R$: a rear-earth ion, $A$: an alkaline-earth ion) and their families.~\cite{Frontera, Tsubouchi, Okimoto1}
In an undoped compound LaCoO$_3$,~\cite{Asai, Tokura} a formal valence of a Co ion is 3+ with a $d^{6}$ electron configuration. 
There are possible three spin states: the LS state with the $(t_{2g})^6(e_g)^0$ configuration, 
the intermediate-spin (IS) state with $(t_{2g})^5(e_g)^1$, and 
the HS state with $(t_{2g})^4(e_g)^2$. 
It is supposed from the electric resistivity and the magnetic susceptibility measurements that the LS band insulator in low temperatures are changed into the HS or IS metallic state with increasing the temperature ($T$).~\cite{Heikes, Raccah, Yamaguchi, Saitoh}
By substitution of $R$ by $A$, corresponding to hole doping into the non-magnetic insulating ground state, 
a system shows ferromagnetic metallic behavior.~\cite{Rao, Rodrigouez, Itoh, Tutsui}
One key point to understand the electronic and magnetic properties in cobaltites are strong correlation between electron conduction and magnetism, i.e. charge and spin degrees of freedom of electrons.~\cite{Suzuki}

Optical irradiation and manipulation in perovskite cobaltites and related materials have been examined by the ultrafast optical pump-probe measurements.~\cite{Iwai2,Okimoto2}
Recently, detailed experiments and analyses have been done in so-called A-site ordered perovskite-type $R$BaCo$_2$O$_{6.\delta}$ crystals
by Okimoto and co-workers.~\cite{Okimoto2}
After pump pulse is introduced into the LS insulator, a metallic state, which is different from the high-temperature metallic state, 
is observed in the optical conductivity spectra.  
This photo-induced state strongly depends on the $R$ species, which is supposed to control a ratio of electron correlation and band width. 
These experiments suggest that strong correlation between electronic and magnetic states remains even in the photo excited state, and  
tell us that the photo-irradiation phenomena in the cobalt oxides should be reexamined from different viewpoint from the photo-induced spin-state change in spin-cross over complexes. 

In this paper, photo-induced spin-state change in correlated electron systems is studied theoretically. 
From the two-orbital Hubbard model, the effective Hamiltonian for the photo-excited state is derived. 
The photo-excited metastable state is obtained through analyses of the effective Hamiltonian by using the exact diagonalization method. 
By irradiation of photons into the LS band insulator near the phase boundary, the HS state is induced. 
It is found that the HS state is stabilized by forming a bound state with a photo-doped hole. 
This bound state brings about a characteristic peak structure in the optical spectra in the photo-excited state. 
A time evolution after photo-irradiation is examined in the time-dependent mean-field scheme. 
A creation of the HS state is caused by a pair annihilation of photo-doped electron and hole. 
This mechanism is able to be confirmed by the time-resolved photoemission spectroscopy experiments. 

In Sect.~II, the model Hamiltonian and the effective model for the photo-excited states are introduced. 
In Sect.~III, numerical results of the electronic states before and after photo-irradiation are presented. 
In Sect.~IV, the time-dependence of the photo-excited states are shown. 
Section~V is devoted to discussion and concluding remarks. 
A brief report for the previous studies in the photo-induced metastable state was published in Ref.~\onlinecite{Kanamori2}. 

\section{model}
\subsection{two-orbital Hubbard model}

We start from the two-orbital Hubbard model as a minimal model to examine the photo-induced spin-state change.
Two orbitals, termed A and B corresponding to the $e_g$ and $t_{2g}$ orbitals in a Co ion, respectively, 
are introduced in each site in a lattice. 
The crystalline field splitting between A and B is represented by $\Delta = \varepsilon _A -
 \varepsilon _B >0$ where $\varepsilon_A$ and $\varepsilon_B$ are the level energies of the A and B orbitals, respectively. 
The model Hamiltonian is given as 
\begin{align}
{\cal H}={\cal H}_U+{\cal H}_t , 
\label{original_H} 
\end{align} 
where we define the on-site term, 
\begin{align}
&{\cal H}_U=\Delta \sum_{i \sigma} c_{i A \sigma}^\dagger c_{i A \sigma}
\nonumber \\
& +U\sum_{i \gamma} n_{i \gamma \uparrow} n_{i \gamma \downarrow}
+U'\sum_{i \sigma \sigma'} n_{i A \sigma} n_{i B \sigma'}
\nonumber \\
& +J\sum_{i \sigma \sigma'}c_{i A \sigma}^\dagger c_{i B \sigma'}^\dagger c_{i A \sigma'} c_{i B \sigma}
  +I\sum_{i \gamma} c_{i \gamma \uparrow}^\dagger c_{i \gamma  \downarrow}^\dagger c_{i \bar{\gamma} \downarrow} c_{i {\bar \gamma}  \uparrow}, 
\label{original_H0}  
\end{align}
and the inter-site term, 
\begin{align}
{\cal H}_t= - \sum_{\langle ij \rangle \gamma \sigma} t_\gamma
  \left ( c_{i \gamma \sigma}^\dagger c_{j \gamma \sigma} +{\rm H.c.} \right ) 
\label{original_Ht} . 
\end{align}
Here, $c_{i \gamma \sigma }^\dagger$ is the electron creation operator at site $i$ with orbital $\gamma (={\rm A, \ B})$ and spin $\sigma (=\uparrow, \downarrow)$.
We define the number operator $n_{i\gamma \sigma }= c_{i\gamma \sigma }^\dag c_{i\gamma \sigma }$ and 
a subscript $\bar{\gamma}=({\rm A, \ B})$ for $\gamma =({\rm B, \ A})$.
The intra-orbital Coulomb interaction $U$, the inter-orbital Coulomb interaction $U'$, the pair-hopping $I$, and the Hund's coupling $J$ are introduced.
The electron transfer integrals between the nearest-neighboring (NN) sites are set to be diagonal with respect to the orbital. 
We assume a relation $t_B < t_A$ by considering 
the transfer-integrals in perovskite oxides, 
and we take $t_A=1$ as a unit of energy.

\begin{figure}[t]
\begin{center}
\includegraphics[width=\columnwidth,clip]{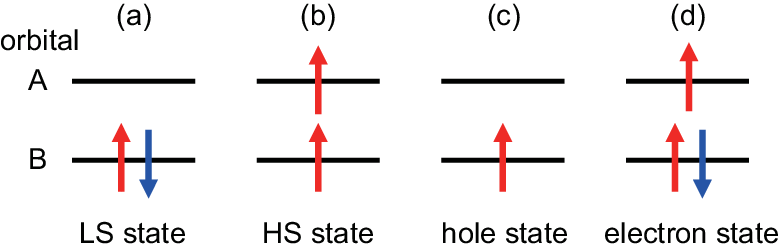}
\end{center}
\caption{
(color online)
Local electronic configurations.  
}
\label{fig:configuration}
\end{figure}
Let us consider the local electronic structure, in which two electrons occupy each site, and the electron transfers are set to be zero. 
The following LS state with $S=0$ and the HS state with $S=1$ are the possible ground states (see Figs.~\ref{fig:configuration}(a) and (b)).  
The eigen function and the eigen energy for the LS state are given as 
\begin{align}
  &\ket{\psi _L }=\left ( f_A c^\dag _{A\uparrow}c^\dag _{A\downarrow} +f_B
  c^\dag _{B\uparrow}c^\dag _{B\downarrow} \right) \ket{0},
\label{eq:sing}
\end{align}
and $E_L=U+\Delta -\sqrt{\Delta ^2 +I^2 }$, respectively, with 
the coefficients  
\begin{align}
 f_B=\left[ 1+\left( \frac{\Delta}{I}- \sqrt{1+ \frac{\Delta^2}{I^2}} \right)^2 \right]^{-1/2}, 
 \label{eq:fb}
\end{align}
and 
\begin{align}
 f_A=\sqrt{ 1-f_B ^2 } . 
 \label{eq:fa}
\end{align} 
The wave functions for the HS state are given by 
\begin{align}
&\ket{\psi _{H+1} }=c^\dag _{A\uparrow}c^\dag _{B\uparrow}\ket{0},
\label{eq:tri+1}\\
&\ket{\psi _{H0} }=\frac{1}{\sqrt{2}}\left ( c^\dag _{A\uparrow}c^\dag
   _{B\downarrow} +c^\dag _{A\downarrow}c^\dag
   _{B\uparrow}\right)\ket{0},
\label{eq:tri0}\\
&\ket{\psi _{H-1} }=c^\dag _{A\downarrow}c^\dag _{B\downarrow}\ket{0},
\label{eq:tri-1}
\end{align}
for $S^z=$+1, 0 and -1, respectively, and the energy in the HS triplet state is $E_H=U'+\Delta -J$.
In the numerical simulations, for simplicity, we assume the relations $U-U'=2J$, $U=4J$ and $I=J$.  
In these assumptions, the LS and HS states are degenerated at $\Delta =J/\sqrt{3}$.

\subsection{Effective Hamiltonian}

\begin{figure}[t]
\begin{center}
\includegraphics[width=\columnwidth,clip]{Fig2.eps}
\end{center}
\caption{(color online)
Adiabatic energy surfaces before and after photo-irradiation. 
Horizontal axis represents a number of the HS state. 
}
\label{fig:schematic}
\end{figure}
One of the main purposes in this paper is to examine a stable steady photo-excited state. 
This is defined as the lowest-energy state inside of the energy surface, where density of the photo-excited electron-hole pairs is fixed. 
A schematic picture is shown in Fig.~\ref{fig:schematic}. 
Bold curves represent the adiabatic energy surfaces before and after photo-irradiation as functions of a number of the HS sites. 
Photons excite a system from the lowest-energy surface to the higher-energy surface. 
Through several kinds of relaxation processes, the system settles down in the lowest energy state in the higher energy surface. 
Instead of time-dependent simulations for the photo-excited dynamics, we examine the lowest-energy state inside of the energy surface where a number of the electron-hole pair is one in a $N$-site system. This state is termed the photo-induced metastable state, from now on. 
We derive the two effective Hamiltonians, where numbers of the electron-hole pairs are zero and one in a $N$-site system. 

The effective Hamiltonians are derived by the perturbational processes from the two-orbital Hubbard model in Eq.~(\ref{original_H}). 
The inter-site transfer term, ${\cal H}_t$, is treated as the perturbation term. 
As for the effective Hamiltonian before the photo-irradiation, the HS and LS states, defined in Eq.~(\ref{eq:sing}) and Eqs.~(\ref{eq:tri+1})-(\ref{eq:tri-1}), respectively, are adopted as the basis states. 
Other local-states, where two electrons occupy each site, have higher energies of the order of $\Delta$, $J$, and $U$ than the LS and HS sates.  
By considering all of the second-order perturbational processes, the Hamiltonian is given as  
\begin{align}
{\cal H}_0
&={\cal H}_U
+J_{HH} \sum _{\left< ij \right>}\left( {\bm   S}_i \cdot {\bm S}_j-1 \right)P^H_i P^H_j 
\nonumber \\
&- J_{LL}  \sum_{\left<ij\right>} P^L_i P^L_j \nonumber\\
&- J_{HL}  \sum_{\left<ij\right>} \left( P^L_i P^H_j+P^H_i P^L_j \right)\nonumber\\
&- J_{++}  \sum_{\left<ij\right>} \left[  I_i ^- I_j ^- \left(  {\bm S}_i \cdot {\bm S}_j-1\right)+\left(  {\bm S}_i \cdot {\bm S}_j-1\right)I_i ^+ I_j ^+   \right]\nonumber\\
&+J_{+-} \sum _{\left< ij\right>}\left[  I_j ^- \left( {\bm S}_i \cdot  {\bm S}_j +1\right) I_i ^+ + I_i ^- \left( {\bm S}_i \cdot
   {\bm S}_j +1\right) I_j ^+  \right] . 
\label{eq:hinitial}
\end{align}
Here, ${\bm S}_i$ is the spin operator defined by ${\bm S}_i=(1/2)\sum _{\gamma \sigma \sigma '}c^\dag_{i\gamma \sigma} {\bm \sigma }_{\sigma \sigma '} c_{i\gamma \sigma '}$ with the Pauli matrices ${\bm \sigma }$, 
and $P^L_i$ and $P^H_i$ are the projection operators for the LS and HS state defined by 
\begin{align}
P_i^L =\ket{\psi _{Li}}\bra{\psi _{Li}}, 
\label{eq:PL}
\end{align}
and 
\begin{align}
P_i^H =\sum_{l=(+, 0, -)}\ket{\psi _{Hli}}\bra{\psi _{Hli}}, 
\label{eq:PH}
\end{align}
respectively. 
The operators $I^+ _i$ and $I^- _i$ change the spin state as 
\begin{align}
I_i^+ =\ket{\psi _{H0i}}\bra{\psi _{Li}}, 
\end{align}
and 
\begin{align}
I_i^- =\ket{\psi _{Li}}\bra{\psi _{H0i}}. 
\end{align}
The prefactors in each term in Eq.~(\ref{eq:hinitial}) are the exchange constants defined by 
\begin{align}
J_{HH}=\frac{t_A ^2+t_B ^2}{U+J} , 
\label{193212_9Aug10}
\end{align}
\begin{align}
J_{LL}=\frac{4f_B ^2 f_A ^2\left( t_A ^2 +t_B ^2 \right)}{2U'+2\Delta_J-U-J},
\label{151643_18Dec10}
\end{align}
\begin{align}
J_{HL}=\left( t_A ^2+t_B ^2 \right)\left ( \frac{f_B^2}{U'+\Delta_J-\Delta } 
+ \frac{f_A ^2}{U'+\Delta_J+\Delta } \right ),
\label{eq:jlh}
\end{align}
\begin{align}
J_{++}=2t_A t_B f_B f_A \left( \frac{1}{U+J}+\frac{1}{2U'-U-J+2\Delta_J} \right),
\label{181622_9Aug10}
\end{align}
\begin{align}
&J_{+-}=2t_A t_B \left( \frac{f_B ^2}{U'+\Delta_J-\Delta }+ 
\frac{f_A ^2}{U'+\Delta_J+\Delta }\right),
\label{181729_9Aug10}
\end{align}
where we define $\Delta_J=\sqrt{\Delta^2+J^2}$. 

The effective Hamiltonian after photo-irradiation is derived in the same way. 
As the unperturbed states, in addition to the LS and HS states, 
we introduce the states where numbers of electrons in a site are one or three (see Figs.~\ref{fig:configuration}(c) and (d)). 
These local states are termed the hole state and the electron state, respectively. 
The wave functions are given as 
\begin{align}   
   \ket{\psi_{h\sigma}}=c^\dag _{A \sigma}\ket{0} , 
\end{align}
and 
\begin{align}
   \ket{\psi_{e\sigma}}=c^\dag _{A \sigma} c^\dag _{B \uparrow} c^\dag _{B \downarrow}\ket{0}, 
\end{align}
respectively. 
The eigen energies are $E_{e}=\Delta +U+2U'-J$ for the electron state and $E_{h}=0$ for the hole state. 
We assume that a number of both the electron state and the hole state is one in a $N$-site cluster. 
The calculated effective Hamiltonian is classified by the electronic states in the NN sites as 
\begin{align}
{\cal H}_1={\widetilde {\cal H}}_{0} +{\cal H}_{eh}+{\cal H}_{e}+{\cal H}_{h}. 
\label{eq:hafter}
\end{align}
The first term ${\widetilde {\cal H}}_0$ corresponds to ${\cal H}_0$ in Eq.~(\ref{eq:hinitial}), where both the electron and hole states are not concerned in the interactions. The second term is for the interactions between the electron state and the hole state. 
The third and fourth terms describe the interactions between the electron state and LS or HS, and the interactions between the hole state and LS or HS, respectively. 
Explicit forms for the Hamiltonian are given in Appendix. 

The ground state before photoirradiation and the photo-induced metastable state are 
obtained in the effective Hamiltonians in Eqs.~(\ref{eq:hinitial}) and (\ref{eq:hafter}), respectively, 
which are analyzed by the exact-diagonalization method based on the Lanczos algorithm. 
Time evolutions in the photo-induced dynamics are calculated in the two-orbital Hubbard model in Eq.~(\ref{original_H}). 

\section{Electronic states before and after photoirradiation}  
\label{sec:exact}
\subsection{Ground State}
\label{sec:GS}

\begin{figure}[t]
\begin{center}
  \includegraphics[width=\columnwidth,clip]{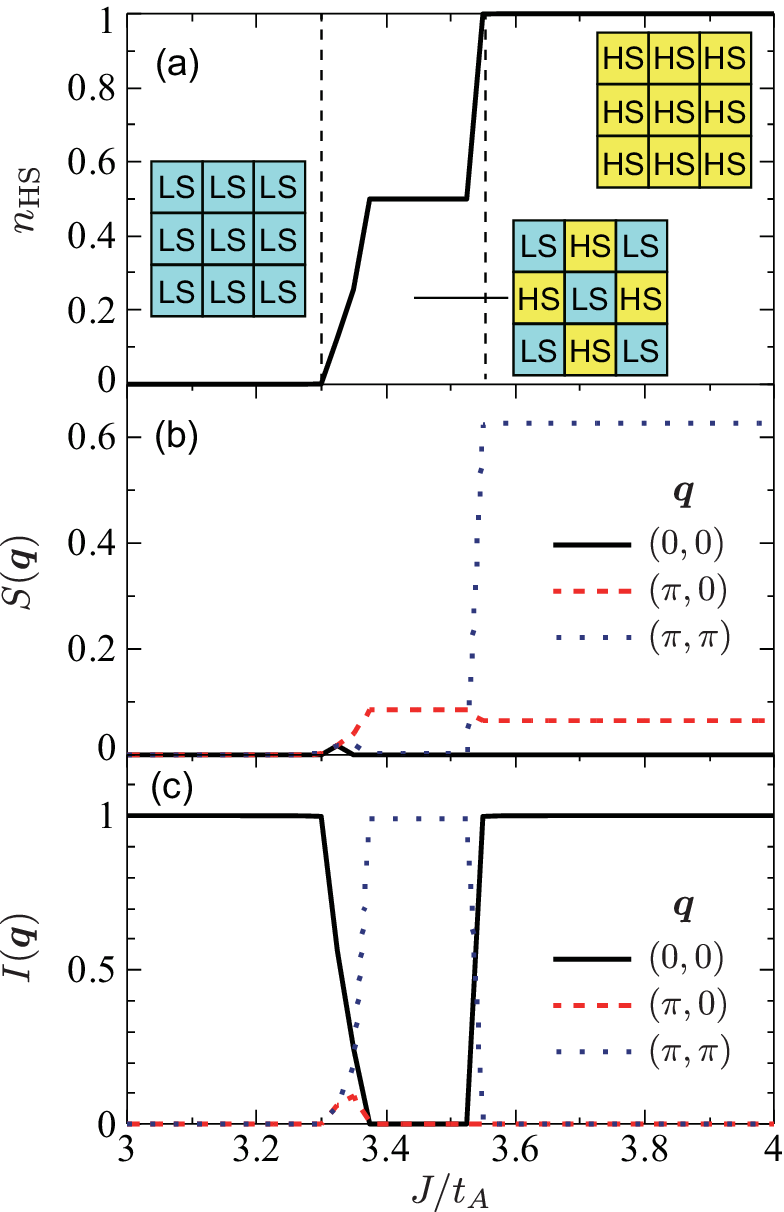}
\end{center}
\caption{
(color online)
(a) The number density of the HS state $n_{\rm HS}$, 
(b) the spin correlation function $S({\bm q})$, 
and (c) the spin-state correlation function $I({\bm q})$ 
in the ground state. 
The parameter values are chosen to be $U=4J$, $U'=2J$, $\Delta =10t_A$, and $t_B=0.05t_A$. 
A two dimensional cluster of the $N=8$ sites with the periodic boundary condition is adopted. 
}
\label{fig:be-stat}
\end{figure}
Electronic structure in the ground state is examined by analyzing the effective Hamiltonian ${\cal H}_0$ in a finite size cluster system. 
Several physical quantities are plotted in Fig.~\ref{fig:be-stat} as a function of the Hund's coupling $J$ at $\Delta/t_A=10$. 
We introduce the number density of the HS states which are estimated from the electron number in the orbital A defined by 
\begin{align}
   n_{HS}=\frac{1}{N}\sum _i \left< n_{iA}\right> ,
\end{align}
the spin correlation function, 
\begin{align}
 &S({\bm q})=\frac{1}{2N^2}\sum _{ij} e^{-i{\bm q}\cdot({\bm r}_i-{\bm r}_j)} \left< {\bm S}_i \cdot {\bm S}_j \right> ,
\label{164230_14Nov10}
\end{align}
and the spin-state correlation function defined by 
\begin{align}
&I({\bm q})=\frac{4}{N^2}\sum _{ij} e^{-i{\bm q}\cdot({\bm r}_i-{\bm r}_j)}   \left< I_i ^z I_j ^z\right> . 
\label{160539_9Aug10}
\end{align}
Here, we define the spin-state operator as a projection operator by 
\begin{align}
I_i ^z=\frac{1}{2}
 \sum _{m=\pm1,0}  \Bigl ( \ket{\psi _{Hmi}}\bra{\psi_{Hmi}}-\ket{\psi _{Li}}\bra{\psi _{Li}} \Bigr ),
\end{align}
which takes $1/2$ and $-1/2$ for the HS and LS states, respectively. 
With increasing $J$, three different phases appear in Fig.~\ref{fig:be-stat}. 
In a region for small $J$, both $n_{HS}$ and $S({\bm q})$ are zero, and $I(0,0)$ is almost one. On the other hand, in a region of large $J$, $n_{HS}$ and $I(0,0)$ are one, and $S(\pi, \pi)$ is the largest. 
Two phases are identified as the LS band insulator and the HS antiferromagnetic Mott insulator. 
Between the two, there is an intermediate phase where $n_{HS}=0.5$, and $I(\pi, \pi)$ is one. These data imply that the HS and LS states are aligned alternately. This phase is termed the spin-state ordered phase.~{\cite{Khomskii}
This alternate ordering of the HS and LS states is caused by the fourth term in the right hand side in Eq.~(\ref{eq:hinitial}); $J_{LH}$ given in Eq.~(\ref{eq:jlh}) represents the attractive interaction between the LS and HS states. 

\begin{figure}[t]
\begin{center}
\includegraphics[width=\columnwidth,clip]{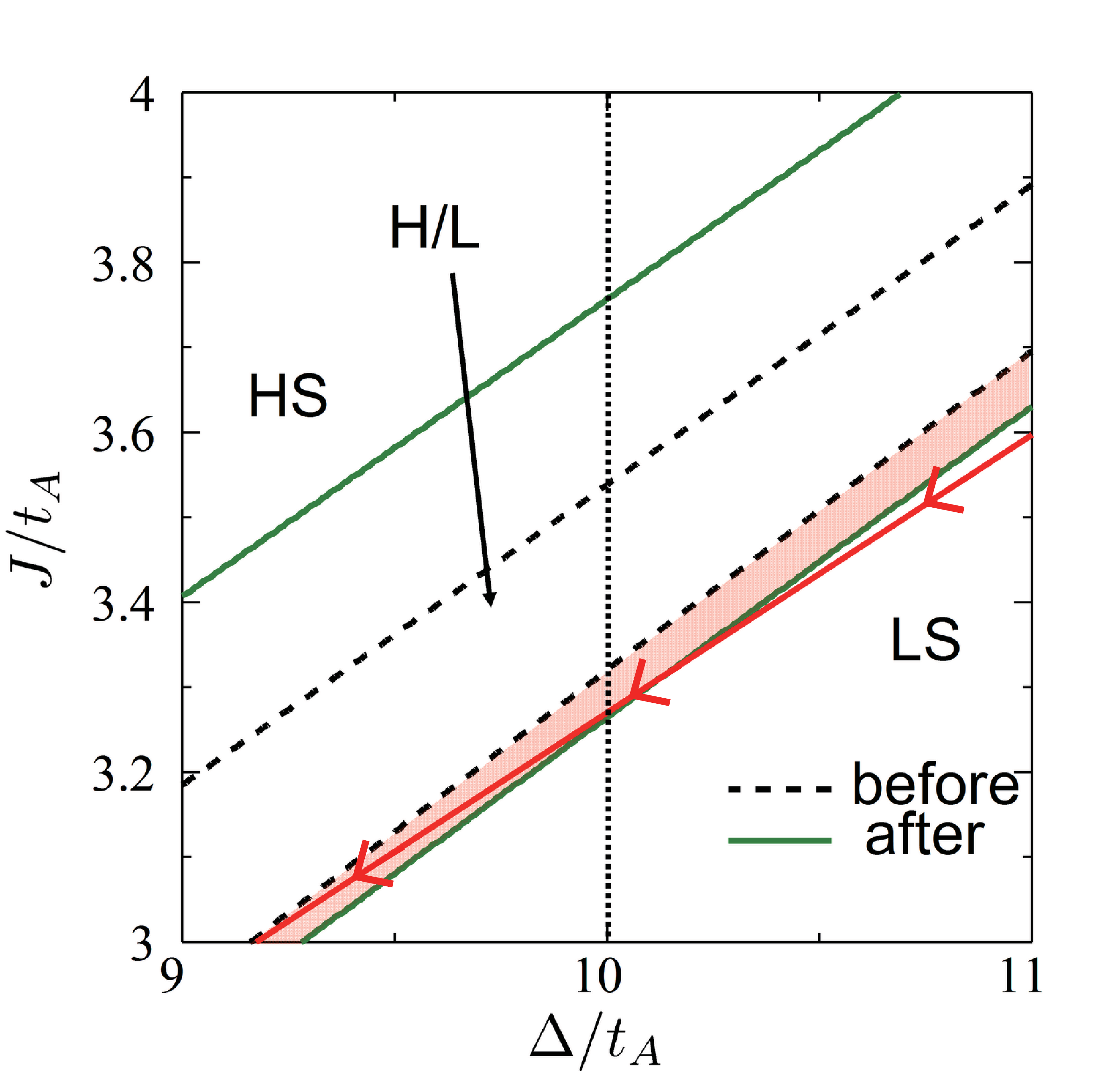}
\end{center}
\caption{
(color online)
Phase diagram in the plane of the crystalline field splitting $\Delta$ and the Hund's coupling $J$. 
Broken and bold lines represent the phase boundaries in the ground state and in the photo-induced metastable state, respectively. 
Abbreviations, HS, LS, and H/L represent the HS phase, the LS phase, and the HS-LS mixed phase, respectively. 
A vertical dotted line represents the parameter region where the data in Figs.~\ref{fig:be-stat} and \ref{fig:af-stat} are calculated. 
Parameter values are chosen to be $U=4J$, $U'=2J$, and $t_B=0.05t_A$. 
A two dimensional cluster of the $N=8$ sites with the periodic boundary condition is adopted. 
}
\label{fig:ph-dig}
\end{figure}
Numerical data in several $J$ and $\Delta$ are summarized in a phase diagram shown in Fig.~\ref{fig:ph-dig}, where the phase boundaries 
in the ground state and those in the photo-induced metastable state are plotted in one figure. 
The detailed results in the metastable state will be presented in Sec.~\ref{Co-rslt-after-State}. 
Here, we identify the LS (HS) phase as a state, where electron numbers of the A orbital is smaller (larger) than 0.3. 
We confirm that the size dependence of the phase boundaries is of the order of $0.01t_A$. 
The LS and HS phases appear in regions of large $\Delta$ and large $J$, respectively.
The spin-state ordered phase appears between the two phases. 
 
\subsection{Photo-induced Metastable State}
\label{Co-rslt-after-State}

\begin{figure}[t]
\begin{center}
\includegraphics[width=\columnwidth,clip]{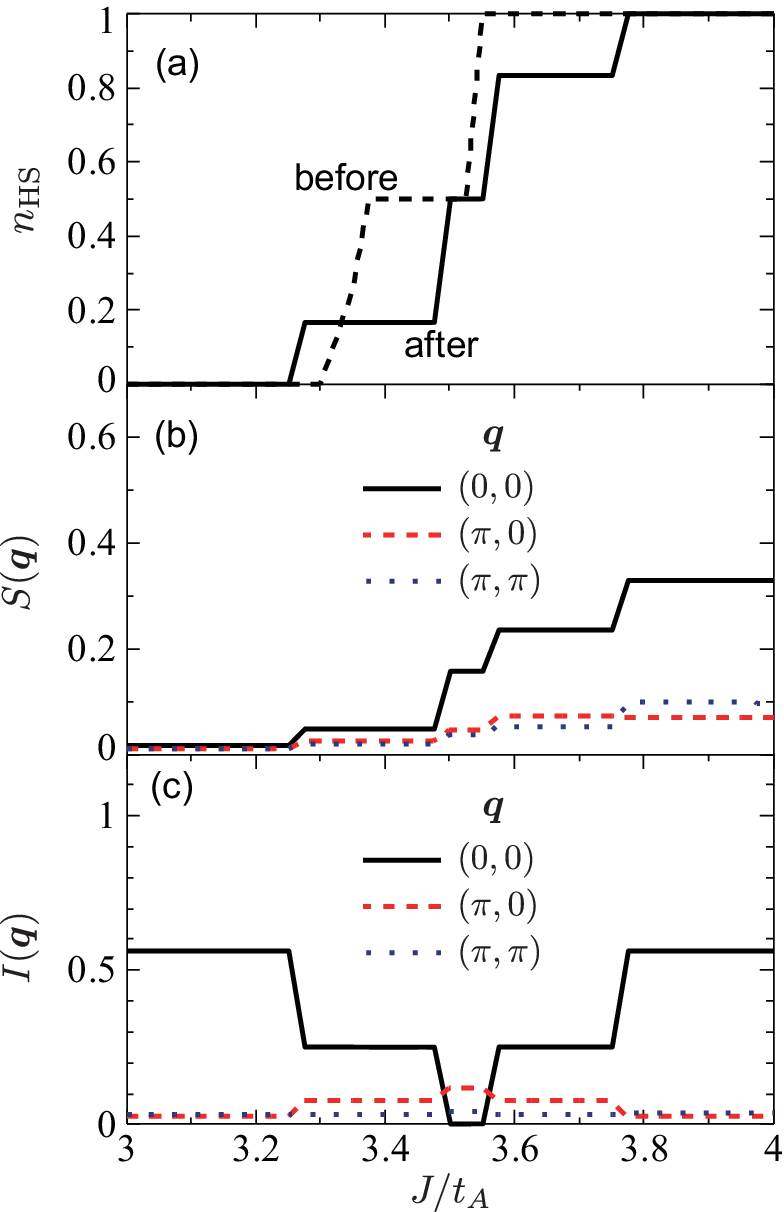}
\end{center}
\caption{
(color online)
(a) The number density of the HS states $n_{\rm HS}$, 
(b) the spin correlation function $S({\bm q})$, 
and (c) the spin-state correlation function $I({\bm q})$ 
in the photo-excited metastable state. 
In comparison, results of $n_{\rm HS}$ in the ground state are also plotted in (a). 
Parameter values are chosen to be $U=4J$, $U'=2J$, $\Delta =10t_A$, and $t_B=0.05t_A$. 
A two dimensional cluster of the $N=8$ with the periodic boundary condition sites is adopted. 
}
\label{fig:af-stat}
\end{figure}

Several physical quantities in the photo-induced metastable state are presented in Fig.~\ref{fig:af-stat} as a function of the Hund's coupling $J$ at $\Delta/t_A=10$. 
The number density of HS states is estimated from a number of electrons in the orbital A defined by 
\begin{align}
    &n_{\rm HS}=\frac{ 1}{N-2}\left( \sum _{i}\left< n_{iA}\right> -1\right) , 
\end{align}
where the electron and hole states are subtracted in a denominator. 
It is shown in the Fig.~\ref{fig:af-stat}(a) that $n_{\rm HS}$ in the photo-induced metastable state is finite between $3.25 < J/t_A <3.30$ where $n_{HS}$ in the ground state is zero. 
Different value between the two states implies that one HS state is generated in the $N$-site cluster. 
This phase in the photo-induced metastable state is distinct from the spin-state ordered phase observed in the ground state; 
the spin-state correlation functions at any ${\bm q}$ are not remarkable, and a weak spin correlation at ${\bm q}=(0,0)$ is observed. 
Detail properties of this phase are introduced latter. 

The phase diagram in the photo-excited metastable state is presented in Fig.~\ref{fig:ph-dig}, together with that in the ground state. 
The phase boundary between the LS and LS-HS mixed phases shifts to a region of the LS phase. 
There is a parameter region where the LS phase in the ground state is changed into the LS-HS mixed phase in the photo-excited metastable state. That is to say, the photo-irradiation induces the HS state in the LS phase at vicinity of the phase boundary. 
We note that the spin-state change also occurs from the HS phase to the mixed phase. 

\begin{figure}[t]
\begin{center}
\includegraphics[width=\columnwidth,clip]{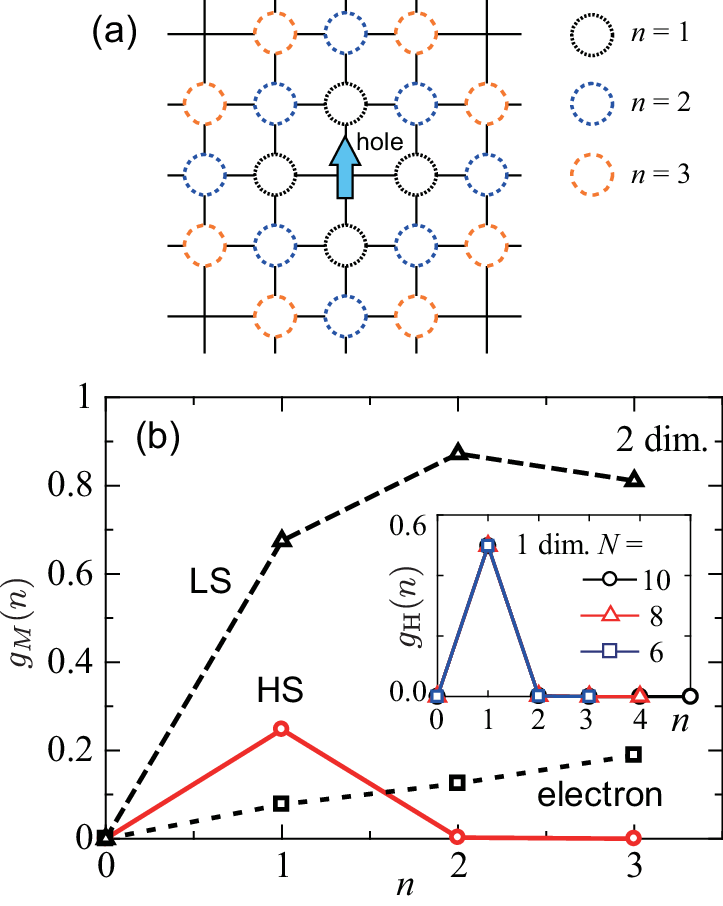}
\end{center}
\caption{
(color online)
(a) A schematic definition of the distribution function $g_M(n)$. 
(b) Distribution function of the electronic states as functions of a distance from the hole state. 
Bold, broken and dotted lines represent the distribution functions for the HS, LS and electron states, respectively. 
Two dimensional $N=10$ site cluster with the periodic boundary condition is adopted. 
The inset shows distribution functions for the HS state in one dimensional $N=$6, 8, and 10 site clusters with the periodic boundary condition. 
Parameter values are chosen to be $J=3.3t_A$, $U=4J$, $U'=2J$, $\Delta =10t_A$, and $t_B=0.05t_A$. 
}
\label{fig:gmj}
\end{figure}
\begin{figure}[]
\begin{center}
\includegraphics[width=\columnwidth,clip]{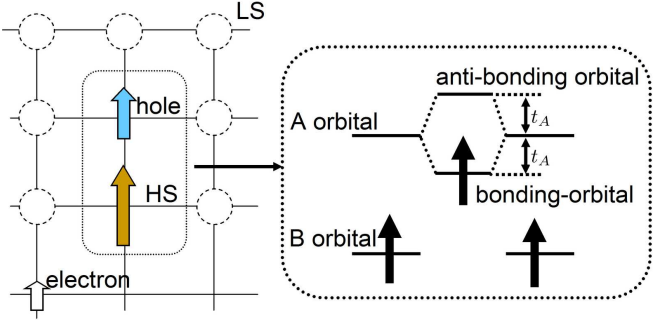}
\end{center}
\caption{
(color online)
(left) A schematic picture of the HS-hole bound state. 
(right) Energy levels inside of the HS-hole bound state. 
}
\label{fig:bonding-schematic}
\end{figure}
Now we examine the electronic structure in the photo-induced HS state in more detail. 
We introduce the electronic-state distribution function defined by 
\begin{align}
    g_M(n)=z_n^{-1} \sum_{j \in nNN} \sum _i \left< P_{i+j}^M  P_i^{ h} \right> , 
\end{align}
where $\sum_{j \in nNN}$ implies a summation of $j$ connecting the $n$-th NN sites of $i$, 
and $z_n$ is a number of the $n$-th NN sites. 
The operator $P^M_i$ $(M=L, H, e)$ is the projection operator for the $M$-state at site $i$. 
The operators for the LS and HS states are define in Eqs.~(\ref{eq:PL}) and (\ref{eq:PH}), respectively, 
and those for the electron and hole states are defined as 
\begin{align}
P_i^e =\sum_{\sigma} \ket{\psi_{e \sigma i}} \bra{\psi_{e \sigma i}}, 
\label{eq:Pe}
\end{align}
and 
\begin{align}
P^h_i =\sum_{\sigma} \ket{\psi_{h \sigma i}}\bra{\psi_{h \sigma i}}, 
\label{eq:Ph}
\end{align}
respectively. This function, $g_M(n)$, describes distribution of the local electronic states at the $n$-th NN sites from the photo-induced hole state, as shown in Fig.~\ref{fig:gmj}(a). 
Numerical results of the distribution functions in a two-dimensional cluster are shown in Fig.~\ref{fig:gmj}(b). 
Parameter values are chosen to be $J=3.3t_A$ and $\Delta =3.3t_A$ in which the HS state is induced by photo-irradiation. 
A characteristic feature is shown in the HS distribution function;  
$g_{H}(n)$ is nearly 0.25 at $n=1$ and zero at $n \ge 2$. 
This implies a local bound state between the HS state and a photo-doped hole state. 
The size dependence of $g_H(n)$ is checked in the one-dimensional clusters, and results are shown in the inset of Fig.~\ref{fig:gmj}(b). 
The HS distribution is located at the NN sites of the hole state, and almost no size dependence is seen in the results. 
Different numerical values of $g_H(n)$ in one and two dimensional clusters, i.e. 0.25 and 0.5, are attributed to difference of $z_n$. 
Spin structure in this bound state is monitored by the correlation function defined by 
$\sum _{i\neq j}\left<  {\bm S}_{j} \cdot {\bm S}_i P_{j}^{H}  P_i^{ h}\right> $
which represents the spin correlation between the hole and HS states. 
Calculated value is about 0.5 which implies a ferromagnetic spin correlation. 
Figure~\ref{fig:gmj} also shows that $g_{e}(n)$ monotonically increases with $n$. 
This is due to the kinetic-energy gain of the photo-excited electron. 
A schematic electronic structure in the photo-induced metastable state is presented in Fig.~\ref{fig:bonding-schematic}. 

Here we discuss a mechanism of the ferromagnetic HS-hole bound state. 
In the ground state, the energy difference per site between the LS state and the HS state is given by 
$\Delta E_{HS-LS}\equiv E_{HS}-E_{LS}=(U'+\Delta-J)-(U+\Delta -\sqrt{\Delta ^2+I^2})$ in the local limit. 
Let us consider a situation that photo-excited electrons and holes are introduced in the LS phase at vicinity of the phase boundary, 
and these photo-carriers move in the system. 
Magnitude of the exchange process between the hole state and the LS state is given by $f_B ^2 t_B$ (see Eq.~(\ref{161539_2Aug10})). 
Thus, when the electron and hole states move around the LS background without generation of the HS state, 
the kinetic energy of the hole state is $-zf_B ^2 t_B$ where $z$ is a number of the NN sites. 
On the other hand, when the ferromagnetic HS-hole bound state is generated, 
one electron occupies the bonding orbital in the bound state and this energy gain is $-t_A$ as shown in Fig.~\ref{fig:bonding-schematic}. 
Since the electron states move in the same ways in both the two cases, energy difference are $\Delta E_{BS}=(-t_A)-(-zf_B^{2}t_B)$. 
When this energy gain due to the bound state, $|\Delta E_{BS}|$, overcomes the energy cost for the HS generation, $\Delta E_{HS-LS}$, the HS-hole bound state is realized. 

\begin{figure}[t]
\begin{center}
\includegraphics[width=\columnwidth,clip]{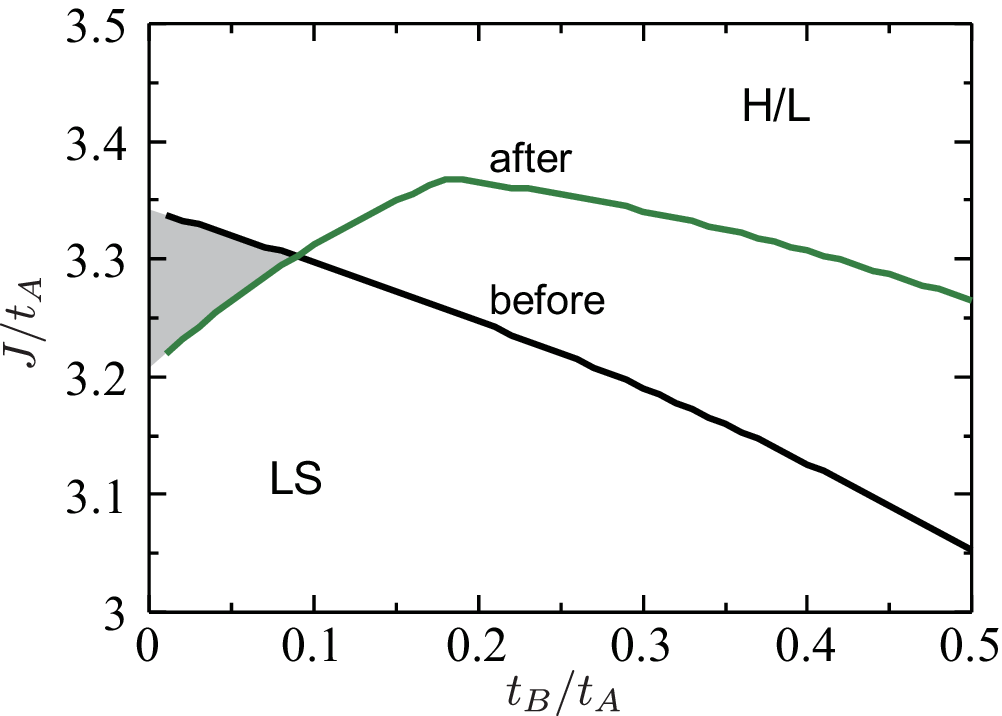}
\end{center}
\caption{
(color online)
Phase diagrams in the ground state and the photo-excited metastable state in the plane of 
$t_B/t_A$ and $J/t_A$.
A shaded area implies a parameter region where the HS state is induced by photo-irradiation.
Two dimensional $N=8$ site clusters with the periodic boundary condition is adopted. 
Parameter values are chosen to be $U=4J$, $U'=2J$, and $\Delta =10t_A$. 
}
\label{fig:pd-tb}
\end{figure}
Above consideration for the energy balance is confirmed in the band width dependence 
of the phase diagram. 
In~Fig.\ref{fig:pd-tb}, the phase diagrams in the ground state and the photo-induced metastable state 
are plotted as functions of a ratio of the band widths for the A and B bands, i.e. $t_B/t_A$. 
In a region of $t_B/t_A < 0.1$, there is a phase space where the LS phase in the ground state is changed into the HS-LS mixed phase in the photo-induced metastable state. 
With increasing $t_B/t_A$, this phase space is shrunken and disappears. 
This is explained from above consideration where stability of the photo-induced HS state is controlled by a factor $\Delta E_{BS}=(-t_A)-(-zf_B^{2}t_B)$. 
This tendency for stability of the HS state is similar to the previous results in the chemical doping.~\cite{Suzuki}

\subsection{Optical Spectra}

In this subsection, we show the optical spectra in the photo-induced metastable state. 
The optical absorption spectra are defined by 
\begin{align}
\alpha ^{\alpha \beta }(\omega )=-\frac{ 1}{N\pi}{\rm Im} 
 \bra{\psi_0}j^{\alpha} \frac{1}{\omega -{\cal H}^{eff}+E_0+i\eta}j^{\beta } \ket{\psi_0}  ,
\label{214254_15Nov10}
\end{align}
where ${\cal H}^{eff}$ is taken to be ${\cal H}_{0}$ in Eq.~(\ref{eq:hinitial}) and ${\cal H}_{1}$ in Eq.~(\ref{eq:hafter}) 
for the ground state and the photo-induced metastable state, respectively, 
$\ket {\psi_0} $ and $E_0$ are the corresponding lowest energy state and energy, respectively,
and $\alpha$($\beta$) represents a Cartesian coordinate.
We introduce the current operator 
\begin{align}
j^\alpha =i\sum _{i \gamma  \sigma }t_\gamma \left (c_{i\gamma \sigma }^\dag c_{i+\alpha \gamma \sigma } -{\rm H.c.} \right ),
\end{align}
which is defined in the restricted Hilbert space in each effective Hamiltonian. 
A damping constant is introduced as $\eta$. 
The optical spectra are calculated by the exact diagonalization method based on the recursion procedure.
Two dimensional finite-size clusters with the periodic boundary condition is adopted. 

\begin{figure}[t]
\begin{center}
\includegraphics[width=\columnwidth,clip]{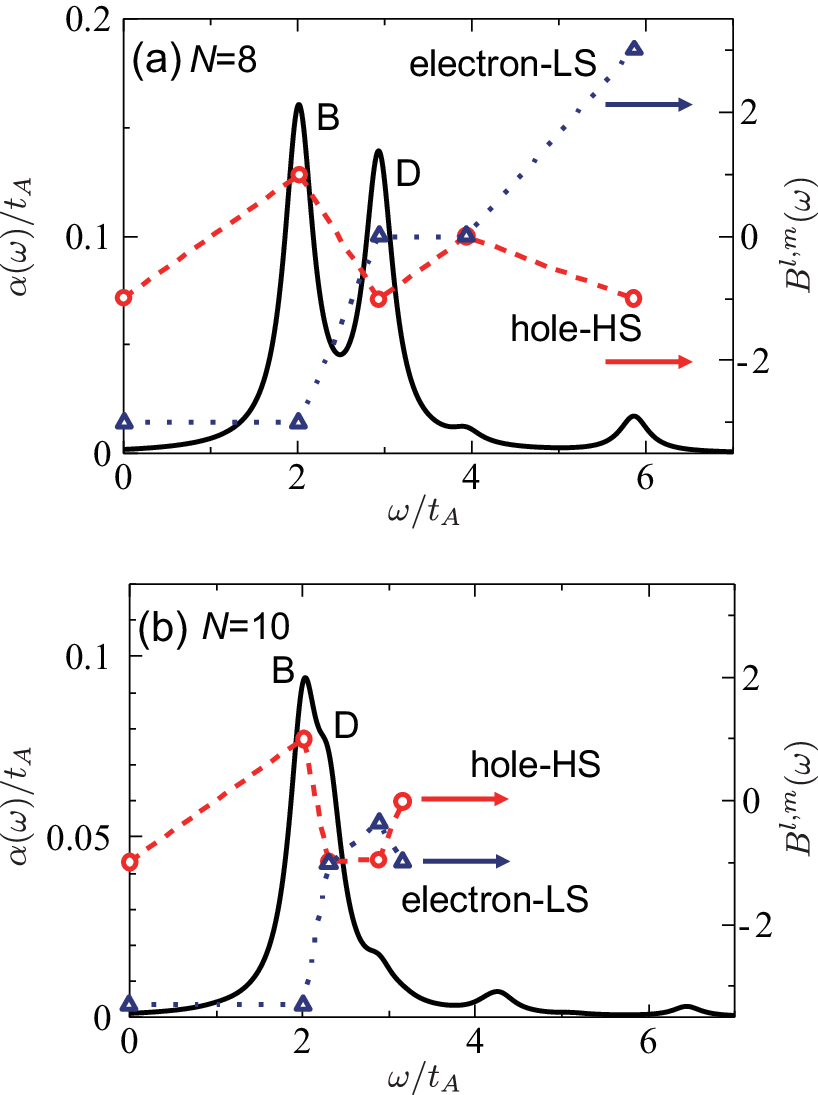}
\end{center}
\caption{
(color online)
Optical absorption spectra in the photo-induced metastable state, where the HS is induced. 
Bond correlation function $B^{(l,m)}(\omega )$ is also plotted. 
Red broken lines and blue dotted lines are for $(l, m)=(h, HS)$ and $(l,m)=(e, LS)$, respectively. 
Cluster size is $N=8$ in (a), and $N=10$ in (b). 
Parameter values are chosen to be $J=3.3t_A$, $U=4J$, $U'=2J$, $\Delta=10.0t_A$, $t_B=0.05t_A$, and $\eta=0.2t_A$.
}
\label{fig:afHL-aw}
\end{figure}
The absorption spectra in the photo-induced metastable state,  
where the HS state is induced by photo-irradiation, 
are shown in Fig.~\ref{fig:afHL-aw}. 
The system size is taken to be $N=8$ and 10. 
Characteristic two peaks appear in the spectra at $\omega =2.1t_A$ and $\omega = 2.8t_A$ in the case of $N=8$. 
These are termed the peaks B and D, and their energies are denoted $\omega_B$ and $\omega_D$, respectively.  
In order to assign these peaks, we calculate the bond correlation function in the excited states given as  
\begin{align}
 B^{(l,m)}(\omega_n )=&-\sum _{\left< ij\right>\gamma \sigma  }  \bra{\psi (\omega_n )} 
 \nonumber \\
 & \times \left ( P_i ^{ m}P_j ^{l}c_{i\gamma \sigma }^\dag c_{j\gamma
    \sigma }P_i ^{l}P_j^{m}+{\rm H.c.} \right)\ket{\psi (\omega_n )}.
\end{align}
where $\ket{\psi(\omega_n)}$ is the eigen function of the Hamiltonian ${\cal H}_1$ corresponding to the final state of the $n$-th optical absorption peak, and $\omega_n$ is its eigen energy. 
Eigen functions and eigen energies are obtained by using the conjugate gradient method. 
This function measures the bond correlation
between the $l$ and $m$ local electronic states in the photo-excited state. 

Numerical results of this correlation function together with the optical absorption spectra are presented in Fig.~\ref{fig:afHL-aw}(a), where we set $(l, m)=(h, HS)$ and $(e, LS)$. 
In the ground state, i.e. $\omega=0$, $B^{(h, HS)}(\omega=0) \sim -1$ and $B^{(e, LS)}(\omega=0) \sim -3$. 
These values are consistent with the picture presented in Fig.~\ref{fig:bonding-schematic}, where 
a photo-doped hole forms a bound state with HS, and a photo-doped electron is located in a bottom of the A-orbital band. 
In the excited state corresponding to the peak B, 
$B^{(e, LS)}(\omega_B) \simeq B^{(e, LS)}(0)$ and 
$B^{(h, HS)}(\omega _B) \sim 1  >  B^{(h, HS)}(0) $. 
This value of $B^{(h, HS)}(\omega _B)$ is interpreted that an electron occupies the antibonding orbital in the HS-hole bound state, and the peak B is assigned as an excitation between the bonding and antibonding orbitals inside of the bound state. 
As for the peak D, $B^{(e, LS)}(\omega _D) > B^{(e, LS)}(0)$ and $B^{(h, HS)}(\omega_D) \simeq B^{(h, HS)}(0)$ which imply that a change in the photo-doped electron motion is concerned in this peak. 

Results in different size cluster of $N=10$ are shown in Fig.~\ref{fig:afHL-aw}(b). 
Numerical values of $B^{(h, HS)}$ and $B^{(e, LS)}$ are almost same with the values in $N=8$. 
Energy of the peak B is almost unchanged, but that of the peak D decreases with increasing $N$. 
These size dependences are consistent with the assignments that the peak B is attributed to the local excitation, and the peak D is related to the kinetic motion of the photo-excited electron. 
We further examine the size dependence of the peak positions in the one-dimensional clusters, and observe that the energy of the peak D decreases with the system size.~\cite{Kanamori2}
This peak is interpreted as a Drude-like component in the thermodynamic limit.

\begin{figure}[t]
\begin{center}
\includegraphics[width=\columnwidth,clip]{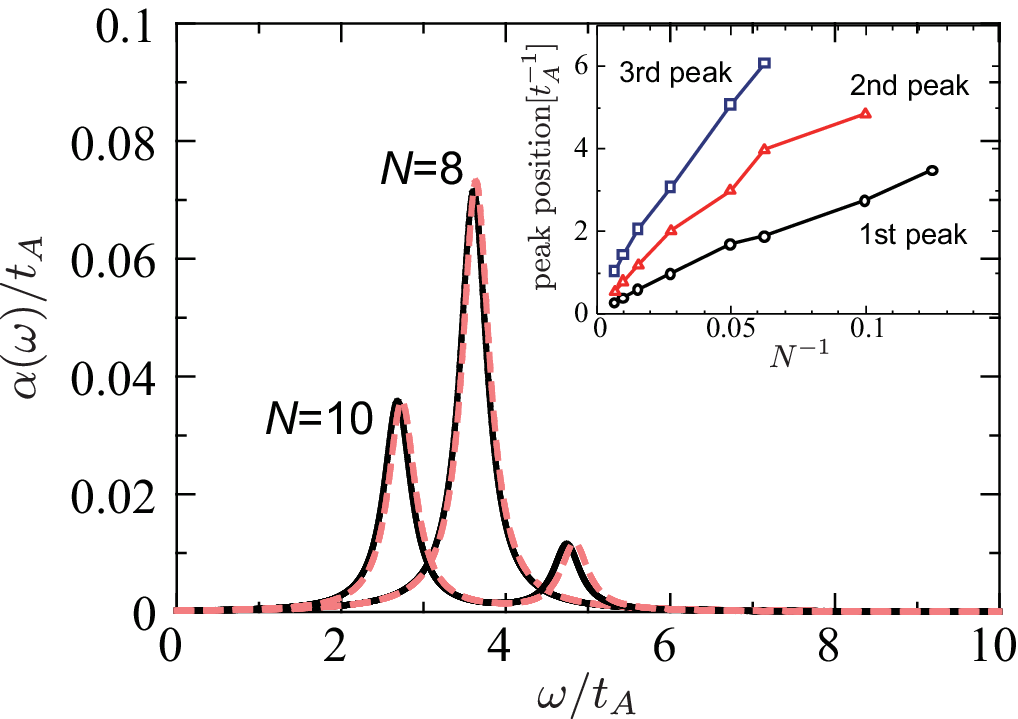}
\end{center}
\caption{
(color online)
Optical absorption spectra in the photo-induced metastable state, where HS state is not induced. 
Broken lines represent the spectra obtained by the HC fermion model. 
Cluster sizes are $N=8$ and $10$. 
Inset shows size dependences of the peak energies in the optical absorption spectra calculated in the HC  fermion model. 
Parameter values are chosen to be $J=3.1t_A$, $U=4J$, $U'=2J$, $\Delta=10.0t_A$,  $t_B=0.05t_A$, and $\eta=0.2t_A$ in the original model, 
and $t_B=0.05t_A, t_{ex}=10^{-10}t_A, \eta=0.2t_A$ in the HC fermion model. 
}
\label{fig:afLS}
\end{figure}
The optical absorption spectra in the photo-induced metastable state, where HS is not induced, are presented in Fig.~\ref{fig:afLS} (see bold lines). 
Two peak structure is observed in $N=10$ and are almost overlapped around $\omega =3.7t_A$ in $N=8$. 
Results are well reproduced by the following hard-core two-fermion model defined by 
\begin{align}
{\cal H}_{HC}&=-t_A\sum _{\left< ij\right>}(a_i ^\dag a_j +{\rm H.c.})-t_B\sum _{\left< ij\right>}(b_i ^\dag b_j +{\rm H.c.})\nonumber\\
    &-t_{ex}\sum _{\left< ij\right>}(a_i ^\dag b_j ^\dag b_i  a_j     +{\rm H.c.}) , 
\label{eq:HHC}
\end{align}
where $a_i$ and $b_i$are the spin-less fermion operators at site $i$ and describe annihilations of the electron state and the hole state, respectively. 
We take a condition of $a_i^\dag b_i^\dag =0$.  
The first and second terms represent kinetic motions of the electron and hole states in the LS phase, respectively, and the third term represents an exchange of the electron and hole states. 
This model is derived in the limiting case of $\Delta >> I$ as follows. 
From Eqs.~(\ref{eq:sing})-(\ref{eq:fa}), we have $\ket{\psi_L }= c^\dag _{B\uparrow}c^\dag _{B\downarrow} \ket{0}$ which is set to be a vacuum, $\ket{\tilde{0}}$, in this model. 
The electron and hole states are defined from this vacuum as $\ket{{\tilde e}}=a_i^\dag \ket{\tilde{0}}$ and 
$|{\tilde h} \rangle=b_i^\dag \ket{\tilde{0}}$, respectively. 
The matrix elements for the exchange of the electron (hole) and LS states, corresponding to the first (second) term in Eq.~(\ref{eq:HHC}),  are given by 
$t_A f_B^2 \sim t_A \ (t_B f_B^2 \sim t_B)$ from Eqs.~(\ref{161539_2Aug10}) and (\ref{161632_2Aug10}). 
The exchange of the electron and hole states, corresponding to the last term in Eq.~(\ref{eq:HHC}), 
are given in the matrix elements in Eqs.~(\ref{160531_2Aug10}) and (\ref{eq:H4}). 
We confirm numerically that this contribution to the optical spectra is much smaller than other terms, and set to be a small constant $t_{ex}=10^{-10}t_A$ in the numerical calculation. 
In this effective model, the current operator along an $\alpha$ direction is given by 
\begin{align}
j_{HC}^{\alpha}=it_A\sum _{i}(a_i ^\dag a_{i+\alpha} -{\rm H.c.})-it_B\sum _{i}(b_i ^\dag b_{i+\alpha} -{\rm  H.c.}),
\end{align}

The optical absorption spectra obtained in the HC model are shown by broken lines in Fig.~\ref{fig:afLS}. 
One pair of the $a$ and $b$ fermions is introduced in the $N$-site clusters. 
Spectra in the effective Hamiltonian are well reproduced by the HC model. 
Size dependences of the peak positions are examined in detail in this model. 
Two dimensional clusters with $N=8$, 10, $4 \times 4$, $6 \times 6$, $8 \times 8$, $10 \times 10$, $12 \times 12$ with the periodic boundary condition are adopted. 
Peak energies for the lowest three peaks are plotted in the inset of Fig.~\ref{fig:afLS} as functions of $N^{-1}$. 
Energies tend to be zero in the thermodynamic limit. 
We interpret that these peaks originate from the metallic behaviors of photo-doped electron and hole.

\section{time-dependence of photo-excited state}
\label{sec:hf}

In this section, we show the real-time evolution of photo-exited state calculated in the mean-field scheme,~\cite{Mclachlan, Terai} and reveal a mechanism of the photo-induced HS state. 

\subsection{Formulation}

Time evolution of the photo-excited state is analyzed. 
A mean-field type decoupling is applied into the Coulomb and exchange interaction terms in the two-orbital Hubbard model in Eq.~(\ref{original_H}) as follows, 
\begin{align}
{\cal H}_{\rm MF}&=
   \sum _{i \gamma  \sigma } n_{i \gamma   \sigma }
   \left[ U\left<n_{i  \gamma  \bar{\sigma }}\right>
   -U'\sum _{\sigma '}\left<  n_{i \bar{\gamma } \sigma '}\right>
   -J\left<n_{i \bar{\gamma } \sigma }\right> \right]\nonumber \\
   &-U\sum _{i \gamma } \left<n_{i  \gamma  \uparrow}\right>\left<n_{i \gamma  \downarrow}\right>
   +U'\sum _{i    \sigma   \sigma ' }\left<n_{i  A \sigma }\right>\left<  n_{i B \sigma '}\right>\nonumber \\
   &+J\sum _{i \sigma } \left<n_{i  A  \sigma }\right>\left<n_{i  B \sigma}\right>
   -\sum _{\left<ij\right> \gamma \sigma}t_\gamma 
   \left(  c_{i \gamma  \sigma }^\dag c_{j \gamma   \sigma }+{\rm H.c.}\right)\nonumber \\
   &+\Delta \sum _{i} n_{i a} , 
\label{174506_18Feb11}
\end{align}
where $\left < \cdots \right >$ implies the average calculated by the time-dependent mean-field wave function,
and a subscript ${\bar \sigma}$ is defined by ${\bar \sigma}=(\uparrow, \downarrow)$ for $\sigma=(\downarrow, \uparrow)$.
We note that the pair-hopping interaction in the Hamiltonian and the Fock terms are not taken into account. 
This is essential to reproduce the electronic states in the case of $t_A=t_B=0$. 
The initial electronic wave function before the photo-excitation is obtained by solving the self-consistent equations. 
The photo-irradiation is simulated by excitations of electrons from the highest occupied levels to the lowest unoccupied ones at time $\tau=0$ with conserving the $z$-component of the total spin-angular momentum and the total momentum. 
The time evolution of the wave function is calculated in the time-dependent mean-field scheme. 
The time-dependent Shr${\rm \ddot{o}}$dinger equation for the $\nu$-th level, $\ket{\phi _\nu (\tau )}$, is given as 
\begin{align}
   &\ket{\phi _\nu (\tau )}=P\exp \left[ -i\int _0 ^\tau d\tau ' {\cal H}_{\rm MF}(\tau' )\right]\ket{\phi _\nu (0)} , 
\label{164222_22Feb11}
\end{align}
where ${\cal H}_{\rm MF}(\tau )$ is the time-dependent Hamiltonian given in Eq.~(\ref{174506_18Feb11}), and $P$ is the time-ordering operator. 
The wave function at time $\tau+d \tau$, where $d \tau$ is short time distance, 
is calculated from the wave function at time $\tau$ by expanding the exponential factor as 
\begin{align}
   &\ket{\phi _\nu (\tau +d\tau )}= \sum _\mu \left< \varphi _\mu
    (\tau)|\phi _\nu (\tau  )\right> e^{ -i \varepsilon _\mu (\tau )d\tau
    }\ket{\varphi _\mu (\tau )},
\end{align}
where $\ket{\varphi _\mu (\tau )}$ is the eigen state of ${\cal  H}_{\rm MF}(\tau )$ with the eigen energy $\varepsilon _\mu (\tau )$. 
In the numerical calculation, we take $d\tau t_A=10^{-3}\sim 10^{-4}$, 
and check that the total energy is conserved within the order of $10^{-2}$ percent. 

\begin{figure}[t]
\begin{center}
\includegraphics[width=\columnwidth,clip]{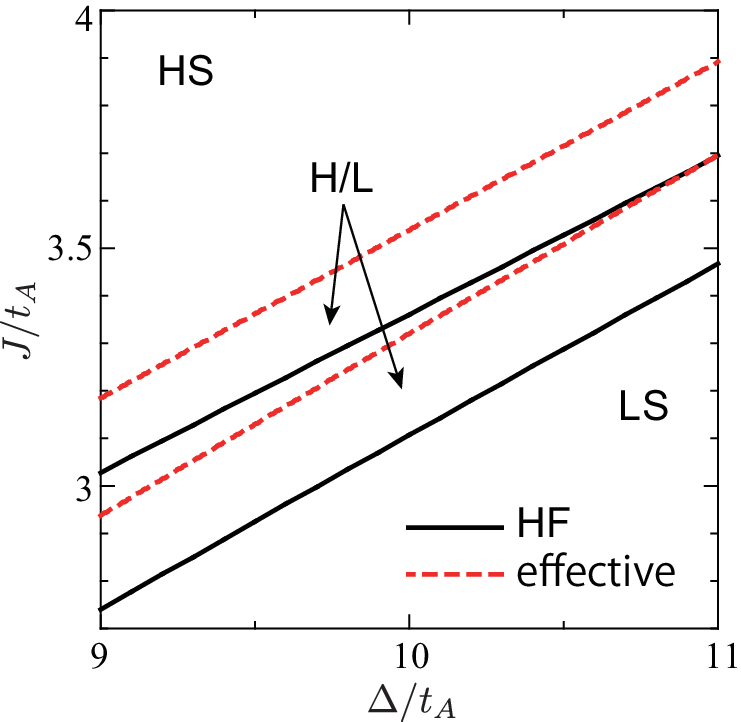}
\end{center}
\caption{
 (color online)
Phase diagram in the ground state obtained by the mean-field approximation. 
Two dimensional $N= 10 \times 10 $ site cluster with the periodic boundary condition is adopted.  
As a comparison, phase boundaries obtained by the exact diagonalization method on the effective Hamiltonian $(N=8)$ are plotted by broken lines. 
Abbreviations, HS, LS, and H/L, represent the HS phase, the LS phase, and the HS-LS mixed phase, respectively. 
Parameter values are chosen to be $U=4J$, $U'=2J$, and $t_B=0.05t_A$. 
}
\label{fig:PD-HF}
\end{figure}
Phase diagram in the ground state is presented in Fig.~\ref{fig:PD-HF}.
Phase boundaries are determined by the HS density.
We also plot the results obtained by the exact diagonalization method applied to the effective Hamiltonian shown in Fig.~\ref{fig:ph-dig}.
Two results are qualitatively similar with each other, although the LS phase 
in the present calculation shifts to a low $J$ region. 

\subsection{Numerical Results}
\begin{figure}[t]
\begin{center}
\includegraphics[width=\columnwidth,clip]{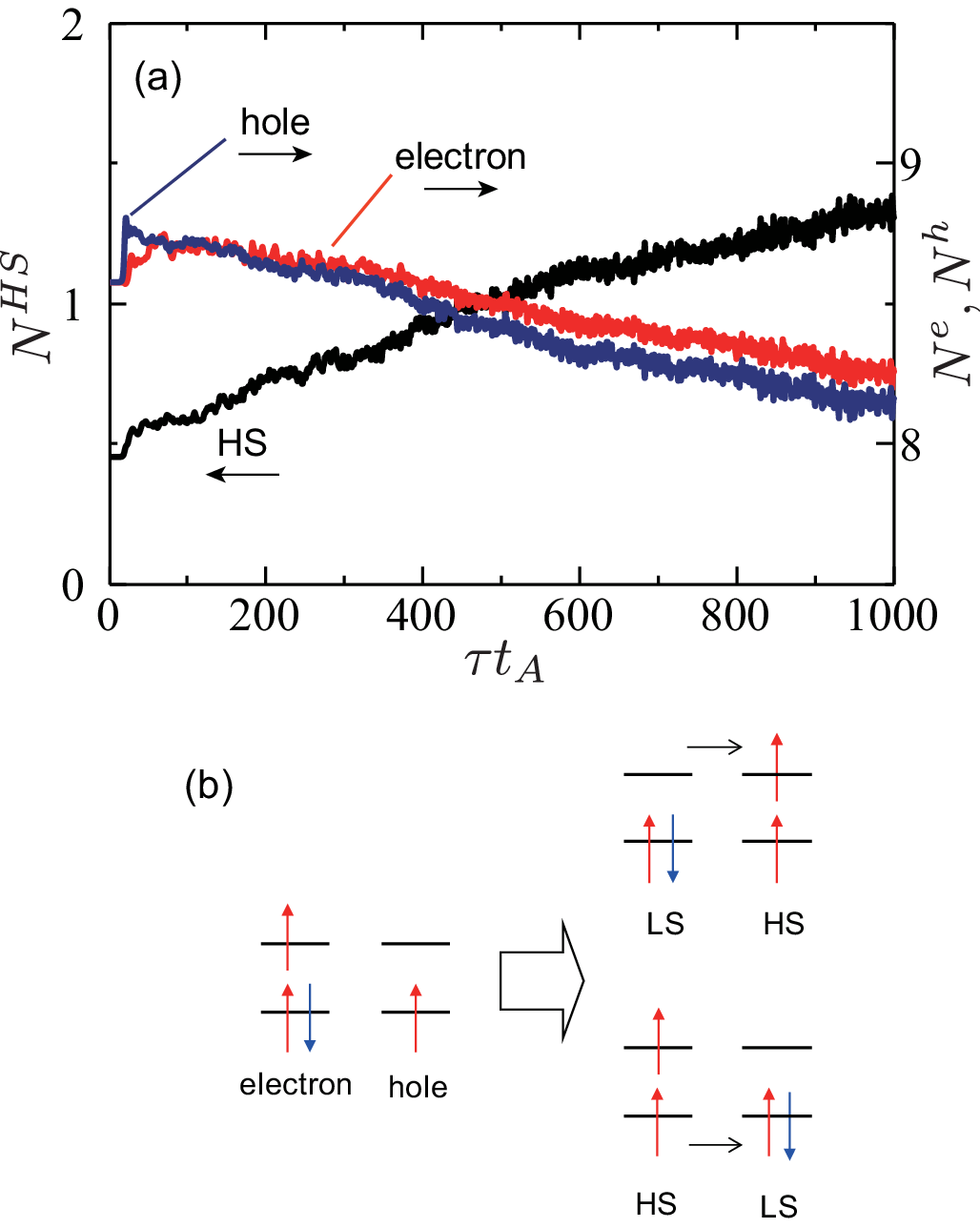}
\end{center}
\caption{
(color online)
(a) Time evolutions of numbers of the HS state, the electron state and the hole state. 
(b) A schematic picture of the electron-hole pair annihilation processes. 
Two dimensional $N= 10 \times 10 $ site cluster with the periodic boundary condition is adopted.  
Parameter values are chosen to be $U=4J$, $U'=2J$, $J=3.1t_A$, $\Delta=10t_A$, $t_B=0.05t_A$, and $N_{ph}=10$. 
}
\label{fig:HF-NHS}
\end{figure}
Time evolution of the photo-excited electronic state is examined in a 
two dimensional $N=10\times 10$ site cluster with the periodic boundary condition. 
A number of photon in the cluster is chosen to be $N_{\rm ph}=10$, which are introduced into the LS phase at a vicinity of the boundary ($J=3.1t_A, \Delta =10t_A$).  
We monitor numbers of the HS state, the photo-doped electron state, and the photo-doped hole state, 
by the following physical quantities, 
$N^{HS}=\sum _{i  \sigma}  N_{i \sigma }^{HS}$, 
$N^{e}=\sum _{i  \sigma} N_{i \sigma } ^{e}$, and 
$N^{h}=\sum _{i  \sigma} N_{i \sigma }^{h}$
with 
\begin{align}
N_{i \sigma } ^{e}=\left<n_{i A \sigma  }\right>(1-\left<n_{i A {\bar \sigma} }\right>)
   \left<n_{i B \sigma  }\right>\left<n_{i B {\bar \sigma} }\right>,
\label{eq:nele}
\end{align}
\begin{align}
N_{i \sigma } ^{h}=(1-\left<n_{i A \sigma }\right>)(1-\left<n_{i A {\bar \sigma} }\right>)
   \left<n_{i B \sigma  }\right>(1-\left<n_{i B {\bar \sigma} }\right>),
\label{eq:nhole}
\end{align}
and 
\begin{align}
N_{i \sigma } ^{HS}=\left<n_{i A \sigma }\right>(1-\left<n_{i A \bar{\sigma}  }\right>)
   \left<n_{i B \sigma  }\right>(1-\left<n_{i B \bar{\sigma} }\right>) ,  
\label{eq:nhs}
\end{align}
respectively. 
It is noticed that these are defined in products of the mean-field number density in each orbital and spin, 
instead of the projection operators such as $P_i^e$ (Eq.~(\ref{eq:Pe})), $P_i^h$ (Eq.~(\ref{eq:Ph})) and $P_i^{H}$ (Eq.~(\ref{eq:PH})), which cannot be calculated directly in the mean-field scheme.  

%
Time dependence of these numbers are plotted in Fig.~\ref{fig:HF-NHS}(a). 
Except for the early time below $\tau t_A=10$, where all three are almost constant,  
$N^{HS}$ increases, and $N^{e}$ and $N^{h}$ decrease monotonically. 
That is, changes in the three numbers are correlated with each other. 
This result is interpreted that the HS states are created by annihilation of the photo-induced electron and hole states. 
Let us consider a situation where the electron and hole states adjoin, as shown in Fig.~\ref{fig:HF-NHS}(b). 
When an electron in the A orbital transfers to the hole state, a LS-HS pair is generated.  
This pair is also generated by the electron transfer in the B orbital.   
This is termed the electron-hole pair annihilation process, from now on. 
  
\begin{figure}[t]
\begin{center}
\includegraphics[width=\columnwidth,clip]{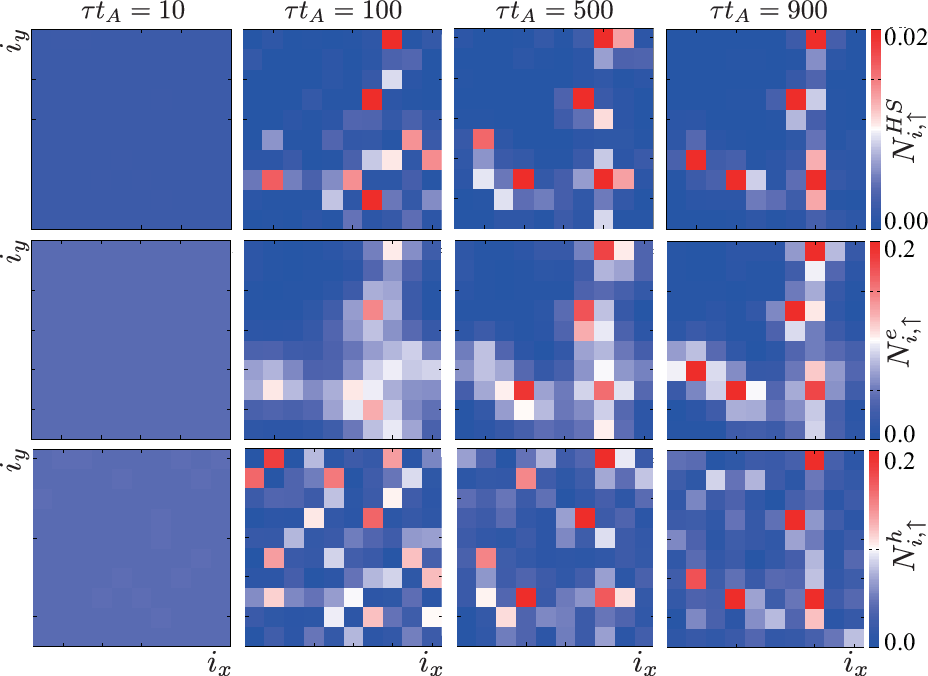}
\end{center}
\caption{
(color online)
Snapshots of the electron state, the hole state and the HS state. 
Time is chosen to be 10$t_A^{-1}$, 100$t_A^{-1}$, 500$t_A^{-1}$ and 900$t_A^{-1}$. 
Two dimensional $N= 10 \times 10 $ site cluster with the periodic boundary condition is adopted.  
Parameter values are chosen to be $U=4J$, $U'=2J$, $J=3.1t_A$, $\Delta=10t_A$, $t_B=0.05t_A$, and $N_{\rm ph}=10$. 
}
\label{fig:HF-CMP}
\end{figure}
Snapshots for the local electronic states are shown in Fig.~\ref{fig:HF-CMP}. 
At $\tau=10t_A^{-1}$, three numbers are almost homogeneous. 
At $\tau=100 t_A^{-1}$, distributions for the electron states start to be inhomogeneous and a vertical shape domain appears in $N^e_{i \uparrow}$. 
At $\tau =500 t_A^{-1}$, the electron states in this vertical-shape domain begin to be localized. 
It is shown that, at the sites where $N^e_{i \uparrow}$ is large, $N^h_{i \uparrow}$ and $N^{HS}$ are also large. 
These data support the mechanism of the HS creation due to the electron-hole annihilation process.  
After the HS states are generated, $N^h_{i \uparrow}$ and $N^e_{\uparrow}$ at the same sites still remain to be large. 
This observation is not contradict to the electron-hole annihilation, 
but is due to the definitions of $N^h_{i \uparrow}$ and $N^e_{i \uparrow}$ [see Eqs.~(\ref{eq:nhole}) and (\ref{eq:nele})]; these are represented in the products of the mean-field number density, instead of the projection operators. 
We have numerically confirmed that, as $N_i^e$ and $N_i^h$ increase, $N_i^{HS}$ increases, and then $N_i^e$ and $N_i^h$ decrease. 
Relation between the numerical results of the real-space snapshots and the HS-hole bound state introduced in the previous section will be discussed in Sect.~\ref{sec:discussion}. 

\begin{figure}[t]
\begin{center}
\includegraphics[width=\columnwidth,clip]{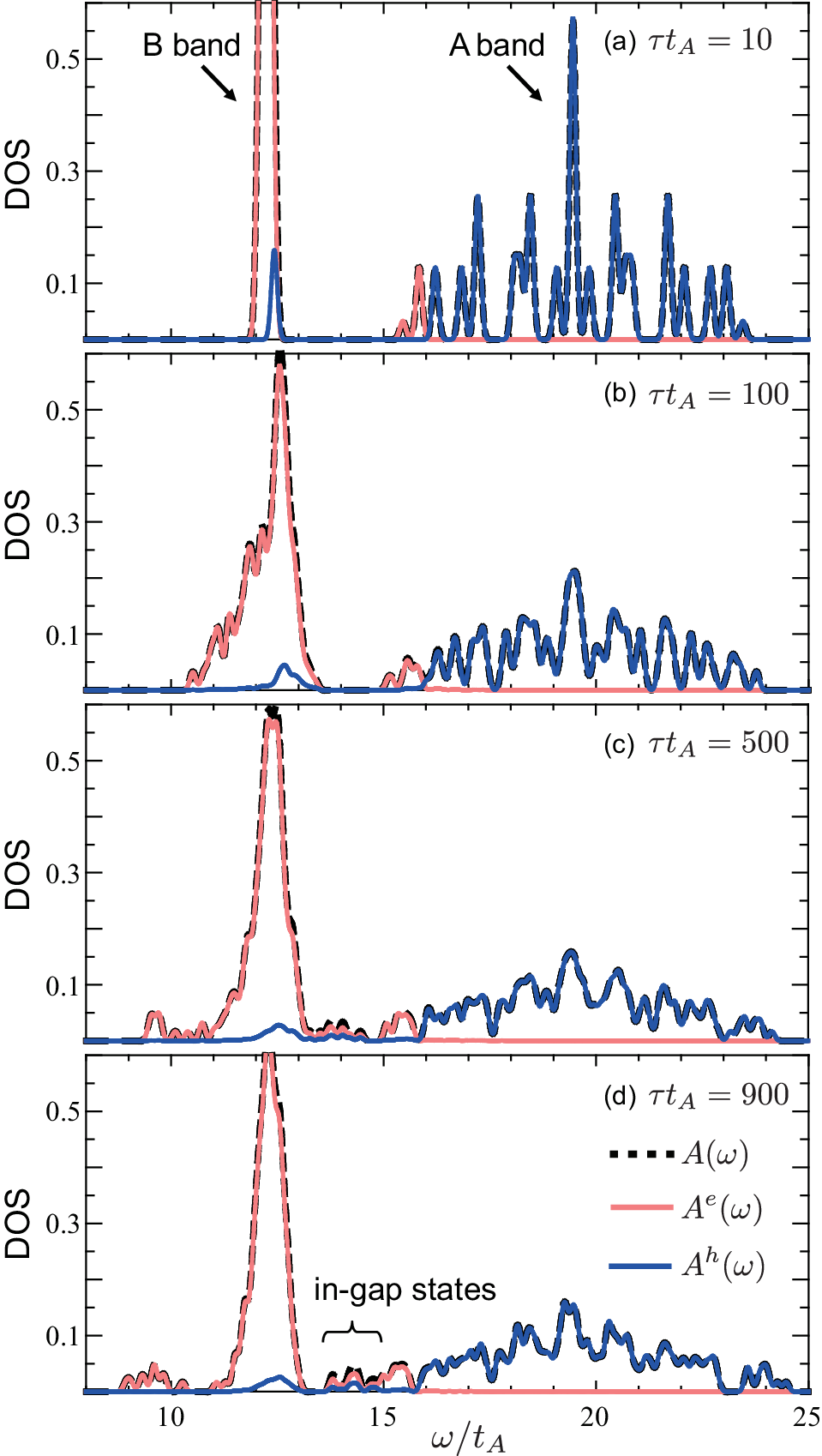}
\end{center}
\caption{
(color online)
Density of states. 
The electron part and the hole part of DOS are represented by 
pink solid and blue solid lines, respectively. 
Time is taken to be 10$t_A^{-1}$, 100$t_A^{-1}$, 500$t_A^{-1}$, and 900$t_A^{-1}$. 
Two dimensional $N= 10 \times 10 $ site cluster with the periodic boundary condition is adopted.  
Parameter values are chosen to be $U=4J$, $U'=2J$, $J=3.1t_A$, $\Delta=10t_A$, $t_B=0.05t_A$, $\eta=0.1t_A$, and $N_{ph}=10$. 
}
\label{fig:HF-DOS}
\end{figure}
The electron-hole pair annihilation processes are also examined by the time dependent density of state (DOS). 
We define DOS as 
\begin{align}
A(\omega )=A^{e}(\omega )+A^{h}(\omega ),
\label{225855_23Nov10}
\end{align}
with the electron part 
\begin{align}
A^{e}(\omega )=\sum _\nu \delta [\omega -\varepsilon _\nu(\tau)] \left< c_\nu^\dag c_\nu \right>,
\label{225903_23Nov10}
\end{align}
and the hole part
\begin{align}
A^{h}(\omega )=\sum _\nu \delta [\omega -\varepsilon _\nu(\tau)] \left< c_\nu c_\nu ^\dag \right>. 
\label{225909_23Nov10}
\end{align}
The operator $c_\nu^\dag$ is the creation operator obtained by diagonalizing the Hamiltonian at time $\tau$, $ \varepsilon _\nu (\tau)$ is the corresponding mean-field energy, and 
$\left<\cdots \right>$ implies the average in terms of the wave function of $\ket{\psi (\tau )}$. 
In the numerical calculation, the delta functions in Eqs.~(\ref{225903_23Nov10}) and (\ref{225909_23Nov10}) are replaced by the Lorentz function with a damping constant $\eta=0.1t_A$. 

Numerical results of the time-dependent DOS are shown in Fig.~\ref{fig:HF-DOS}. 
At $\tau=10 t_A^{-1}$, an energy gap exists between the narrow B band and the wide A band. 
Tiny weights of the hole and electron parts of DOS are observed in the top of the B band and the bottom of the A band, respectively. 
At time $\tau= 500 t_A^{-1}$, the top of the B band and the bottom of the A band start to separate from the main bands. 
Finally, at $\tau=900 t_A^{-1}$, the original gap is almost filled out by in-gap states.
These data are consistent with the localization of the electron and hole states observed in the snapshots in Fig.~\ref{fig:HF-CMP}. 

Based on these results, we consider the energy balance in the electron-hole pair annihilation processes. 
On-site mean-field energies of the LS, HS, electron, and hole states are given by 
$E^{LS}_{\rm MF}=U$, $E^{HS}_{\rm MF}=U'+\Delta -J$,
$E^{e}_{\rm MF}=U+2U'+\Delta -J$, and $E^{h}_{\rm MF}=0$, respectively. 
When one electron-hole pair is changed into one LS state and one HS state, 
the on-site energy is changed as   
$\Delta E_{eh \to LH}
\equiv \bigl(E^{LS}_{\rm MF}+E^{HS}_{\rm MF}\bigr)-
 \bigl(E^{e}_{\rm MF}+E^{h}_{\rm MF} \bigr )=-U'$ 
which is negative, i.e. energy loss. 
This energy is compensated by the kinetic energy of the hole and electron states, which are not concerned in the pair-annihilation processes.
This is confirmed in DOS at $\tau t_A=900$ (see Fig.~\ref{fig:HF-DOS}(d)); 
the hole part of DOS in the B-orbital band distributes not only to the top of the band, but also down to the middle of the band. 
This indicates increasing of the kinetic energy of holes with time.

In the last part of this section, we examine, on the time-evolution of the photo-induced HS generation, roles of the relativistic spin-orbit (SO) interaction which breaks the spin angular-momentum conservation.
Here we mimic the SO interaction in the $3d$ orbitals as follows, 
\begin{align}
{\cal H}_{SO}=i\xi \sum _i \left (
    c_{i a \uparrow}^\dag c_{i b \downarrow}
   +c_{i a \downarrow}^\dag c_{i b \uparrow}
   -c_{i b \uparrow}^\dag c_{i a \downarrow}
   -c_{i b \downarrow}^\dag c_{i a \uparrow}
   \right ), 
   \label{142227_22Nov10}
\end{align}
with the SO interaction constant $\xi$. 
It is demonstrated that, when this interaction acts on the LS state, the HS state is created as follows, 
\begin{align}
   &{\cal H}_{SO}c_{i b \uparrow}^\dag c_{i b \downarrow}^\dag\ket{0}
   =-i\xi \left(  c_{i a \uparrow}^\dag c_{i b \uparrow}^\dag
   -c_{i a \downarrow}^\dag c_{i b \downarrow}^\dag \right ) \ket{0} . 
\end{align}
  
\begin{figure}[t]
\begin{center}
\includegraphics[width=\columnwidth,clip]{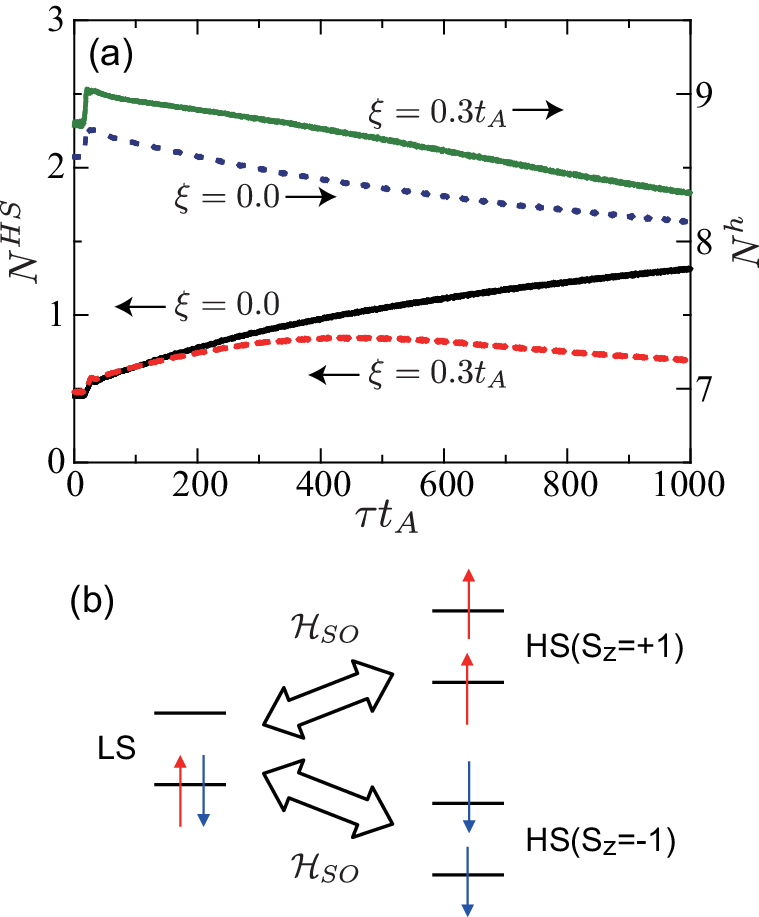}
\end{center}
\caption{
(color online)
(a) Numbers of the HS state, the electron state and the hole state in the model with the spin-orbit interaction. 
Two dimensional $N= 10 \times 10 $ site cluster with the periodic boundary condition is adopted.  
Parameter values are chosen to be $U=4J$, $U'=2J$, $J=3.1t_A$, $\Delta=10t_A$, $t_B=0.05t_A$, and $N_{ph}=10$. 
Ten data sets with different initial values for the time evolutions are averaged. 
(b) A schematic picture of the transition between the HS state and the LS state due to the SO interaction. 
}
\label{fig:SOint_HS}
\end{figure}
Numerical results of the time evolutions of $N^{HS}$ and $N^{h}$ are presented in Fig.~\ref{fig:SOint_HS}, 
where the SO interaction constant is taken to be $ \xi =0$ and $0.3t_A$. 
Before $\tau =200 t_A^{-1}$, the SO interaction effects are not seen in $N^{HS}$. 
However, beyond $\tau  =200 t_A^{-1}$, $N^{HS}$ starts to decrease in the case of a finite $\xi$. 
The observed reduction of the HS state in the case of finite $\xi$ is due to the transition from the HS to LS states through the SO interaction, as shown schematically in Fig.~\ref{fig:SOint_HS}(b). 
This result indicate that roles of the SO interaction on the spin-state transition is destructive rather than constructive.

\section{Discussion and Conclusion}
\label{sec:discussion}

In this section, we remark i) a connection between the calculated results in the photo-excited metastable state and the time dependent simulation in the photo-excited state, which are presented in Sects.~\ref{sec:exact} and \ref{sec:hf}, respectively, and ii) implications of the present theoretical results to the recent experiments. 

\begin{figure}[t]
\begin{center}
\includegraphics[width=\columnwidth,clip]{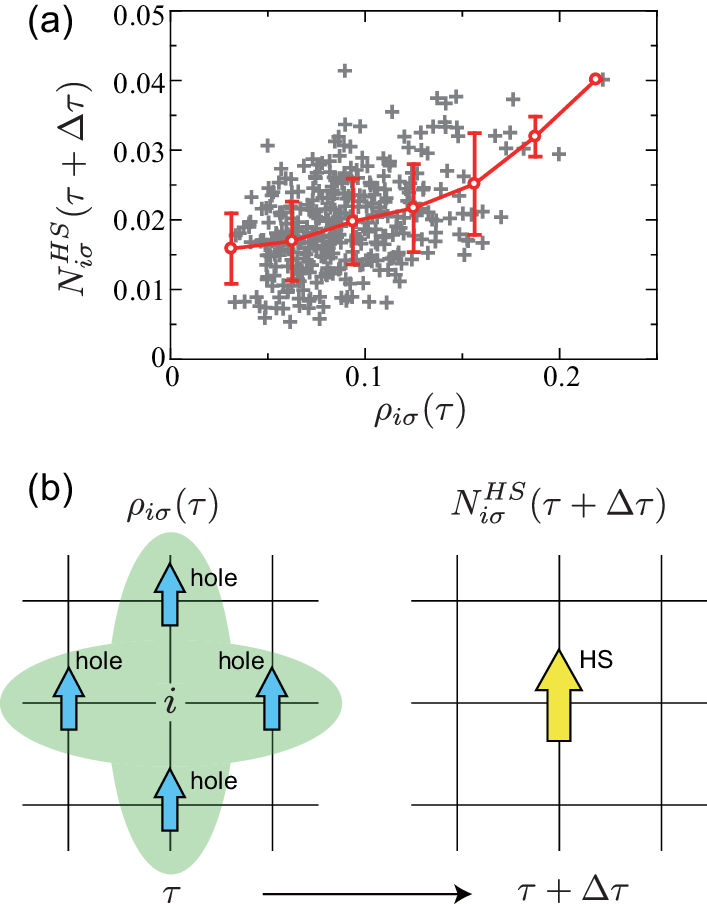}
\end{center}
\caption{
(color online)
Correlation between the number density of the local HS state at time $\tau$, $N_{i \sigma}^{HS}(\tau+\Delta \tau)$, and the number density of the hole state around the site $i$ at time $\tau'$,
$\rho_{i \sigma}(\tau)$. We take $\Delta \tau=10/t_A$. 
Two dimensional $N= 10 \times 10 $ site cluster with the periodic boundary condition is adopted.  
Parameter values are chosen to be $U=4J$, $U'=2J$, $J=3.1t_A$, $\Delta=10t_A$, $t_B=0.05t_A$, $N_{ph}=10$, and 
$\Delta \tau=10t_{A}^{-1}$.}
\label{fig:corre}
\end{figure}
In Sect.~\ref{sec:hf}, we show in the time-dependent simulation that a pair annihilation of photo-doped electron and hole generates a HS state. In this scheme, here we discuss a stability of the HS state and a role of the photo-doped hole. 
A correlation between the local HS state and the hole state around the local HS state is examined numerically. 
We introduce the number density of the local HS state at time $\tau$, $N_{i \sigma}^{HS}(\tau)$, defined in Eq.~(\ref{eq:nhs}), and 
the number density of the hole state around the site $i$ at time $\tau'$ denoted as 
$\rho_{i \sigma}(\tau') \equiv \sum_{j}' N_{j \sigma}^{h}(\tau')$. 
We define $N_{i \sigma}^{h}(\tau')$ in Eq.~(\ref{eq:nhole}), and  introduce a symbol $\sum_{j}'$ which implies a summation for the NN sites of $i$. 
Data sets of $\rho_{i \sigma}(\tau)$ and $N_{i \sigma}^{HS}(\tau+\Delta \tau)$ with $\Delta \tau=10t_A^{-1}$ are obtained in 100 times simulations with different initial states. 
Numerical results are shown in Fig.~\ref{fig:corre}. 
A positive correlation between the two quantities is seen in this figure. 
In particular, in a region of high density of hole, $(\rho_{i \sigma}(\tau) \gtrsim 0.1)$, a number of data for small $N_{i \sigma}^{HS}(\tau+\Delta \tau)$ is a few. 
On the other hand, in a region of low density of hole, $(\rho_{i \sigma}(\tau) \lesssim 0.1)$, value of $N_{i \sigma}^{HS}(\tau+\Delta \tau)$ distributes. 
These results are interpreted that in the case that the hole density around the photo-induced HS state is low, a probability of the survival HS state number is randomly distributed. 
On the other hand, in the case of high hole density around the HS state, the HS density increases with increasing the hole density. 
These relations between the photo-doped hole and the HS state are consistent with the results in Sect.~\ref{sec:exact}, where the HS state is stabilized by forming the HS-hole bound state.  

Next we compare the present calculated results with the experimental data reported in Ref.~\onlinecite{Okimoto2}. 
As introduced previously, key points in the optical pump-probe experiments in $R$BaCo$_2$O$_{6.\delta}$
are a) a photo-induced metallic state is different from the high-temperature metallic state, 
and b) this photo-induced state strongly depends on the $R$ species. 
From the calculated results, we propose that the observed metallic state is attributed to the HS-hole bound state. 
Experimental spectral weight induced by the photon pumping is interpreted to be the dipole transition inside of the bound state. 
We have checked by the exact diagonalization method in a small size cluster that a clear bound state between thermal hole carriers and the HS states is not stabilized in finite temperatures.~\cite{Kanamori2}
It is well known that, in the perovskite crystal, the electron transfer intensity is systematically controlled by the $R$ species through a changing of a Co-O-Co bond angle.
The smaller the ionic radius of the $R$ ion is, the smaller the $e_g$ band width is. 
With increasing the ionic radius from Tb to Sm, the photo-induced metallic state is remarkably seen in the experimental optical conductivity spectra. 
These data correspond to the calculated results in the phase diagram in Fig.~\ref{fig:ph-dig}. 
Increasing of the transfer integral of the A band, being equivalent to decreasing of $\Delta/t_A$,
is indicated by the arrow in this phase diagram. 
A system is transferred from the phase, where the spin-state is not changed by photo-excitation, to 
the phase, where the HS state is induced by photo-irradiation. 
This consistency between the theory and the experiments is additional evidence of existence of the photo-induced HS-hole bound state. 

In conclusion, we study the photo-induced spin-state change in a correlated electron system. 
The photo-induced metastable state is examined in the effective Hamiltonian which is derived by the two-orbital Hubbard model. 
By photo-irradiation into the LS phase near the phase boundary with the mixed phase, the HS state is induced and is stabilized by forming a bound state with a photo-doped hole state. 
The optical transition inside of this bound state appears. 
Time dependent simulation for the photo-excited state is also performed on the two-orbital Hubbard model in the time-dependent mean-field scheme. 
A pair annihilation of the photo-doped electron and hole states generates the HS state. 
This process reflects on the time-dependent DOS. 
The present results propose a new state of the photo-excited matter in correlated electron system with multi-degrees of freedom.

\begin{acknowledgments}
Authors would like to thank H. Matsueda, Y. Inoue, Y. Okimoto, S. Koshihara, S. Iwai and T. Arima for their valuable discussions.
This work was supported by KAKENHI from MEXT, Optical Science of Dynamically Correlated Electrons (DYCE), 
Tohoku University "Evolution" program, and Grand Challenges in Next-Generation Integrated Nanoscience.
YK is supported by the global COE program "Weaving Science Web beyond Particle-Matter Hierarchy" of MEXT, Japan.
Parts of the numerical calculations have been performed in the supercomputing systems in ISSP, University of Tokyo, and Kyoto University. 
\end{acknowledgments}

\appendix
\section{Effective Hamiltonian for the Photo-excited Meta-stable State}
 \label{Hpd-detail}

In this Appendix, explicit formulae of the effective Hamiltonian for the photo-induced metastable state are presented. 
Matrix elements in terms of the electronic states in NN sites are shown. 
These are classified by the electron number, $n$, and the $z$-component of the total spin-angular momentum, $S_z$, as ${\cal H}^{(n, S_z)}$. 
The wave functions in the two sites are denoted as $\ket{\psi_i, \psi_j}$. 
In the following notation, each term in the Hamiltonian in Eq.~(\ref{eq:hafter}) is given as 
${\cal H}_{eh}={\cal H}^{(4,0)}+{\cal H}^{(4,1)}$, ${\cal H}_{e}={\cal H}^{(5,3/2)}+{\cal H}^{(5,1/2)}$ 
and ${\cal H}_h={\cal H}^{(3,3/2)}+{\cal H}^{(3,1/2)}$. 

(1) $(n=3,\ S_z=3/2)$
\begin{align}
 {\cal H} ^{(3,3/2)}=\ 
\begin{pmatrix}
0 &t_A \cr
t_A &0
\end{pmatrix}   . 
\label{144407_2Aug10}
\end{align}
The basis set is $\{ \ket{\psi_{h \uparrow}, \psi_{H+1}}  \ket{\psi_{H+1}, \psi_{H \uparrow}} \} $. 

(2) $(n=3,\ S_z=1/2)$ 
\begin{align}
 {\cal H}^{(3,1/2)}=
\begin{pmatrix} 
   -J_{hH}         &\frac{J_{hH}}{\sqrt{2}} & 0                & \frac{t_A}{\sqrt{2}}    & 0     & 0        \cr
   \sqrt{2}J_{hH} &-\frac{J_{hH}}{2}       & \frac{t_A}{\sqrt{2}}     & \frac{t_A}{2}  & 0     & 0         \cr
    0               & \frac{t_A}{\sqrt{2}} &-J_{hH}          &\frac{J_{hH}}{\sqrt{2}}& 0   & 0             \cr
    \frac{t_A}{\sqrt{2}}  & \frac{t_A}{2}  &\sqrt{2}J_{hH}  &-\frac{J_{hH}}{2}       & 0      & 0     \cr
    0               & 0               & 0                & 0           &-J_{hL}& f_B ^2 t_B      \cr
    0               & 0               & 0                & 0            & f_B ^2 t_B      &-J_{hL} 
\end{pmatrix}.
\label{161539_2Aug10}
\end{align}
where 
\begin{align}
J_{hH}&=\frac{t_A ^2}{4J}+\frac{t_B^2}{U+U'} \nonumber \\
&+\frac{t_B^2 f_B ^2}{\Delta+J+U-U'-\Delta_J } \nonumber\\
&+\frac{t_B^2 g_B ^2}{\Delta+J+U-U'+\Delta_J} , 
\label{141915_2Aug10}
\end{align} 
\begin{align} 
J_{hL}
&= \frac{t_B^2 f_B^2 g_B^2}{2 \Delta_J} +\frac{t_B^2 f_A^2}{\Delta +2U'-J+\Delta_J }        \nonumber \\
&+\frac{3f_A ^2 t_A ^2}{2}  \frac{1}{\Delta-U+U'+\Delta_J-J}   \nonumber \\
&+\frac{3f_A ^2 t_A ^2}{2}  \frac{1}{\Delta-U+U'+\Delta_J+J}  , 
\end{align}
with 
\begin{align}
g_B=\left[ 1+\left( \frac{\Delta}{I}+ \sqrt{1+ \frac{\Delta^2}{I^2}} \right)^2 \right]^{-1/2} . 
\end{align}
The basis set is 
$\{ \ket{ \psi_{h \downarrow},\psi_{H+1}}, \ket{\psi_{h \uparrow}, \psi_{H0}}, \ket{\psi_{H+1}, \psi_{h \downarrow}},$  
$\ket{\psi_{H0}, \psi_{h \uparrow}}, \ket{\psi_{h \uparrow}, \psi_{L}}, \ket{ \psi_{L}, \psi_{h \uparrow}} \} $. 

(3) $(n=4,\ S_z=1)$
\begin{align}
   {\cal H} ^{(4,1)}=
\begin{pmatrix}
    J_{eh1} & \alpha _{eh} J_{eh1} \cr
    \alpha _{eh} J_{eh1} &J_{eh1} 
\end{pmatrix} , 
\label{160531_2Aug10}
\end{align}  
where
\begin{align}
&J_{eh1}=\left( t_A ^2 +t_B ^2 \right)
\Bigl ( \frac{f_B^2}{U'-\Delta +\Delta_J}+\frac{g_B ^2}{U'-\Delta -\Delta_J} \Bigr ),
\label{142739_2Aug10}
\end{align}
and 
\begin{align}
\alpha _{eh}= \frac{2 t_A  t_B }{ t_A ^2 +t_B ^2 } . 
\end{align}
The basis set is 
$\{ \ket{\psi_{e \uparrow}, \psi_{h \uparrow}}  \ket{\psi_{h \uparrow}, \psi_{e \uparrow}} \}$. 

(4) $(n=4,\ S_z=0)$
\begin{align}
{\cal H}^{(4,0)}=\frac{1}{2}
&\begin{pmatrix}
   J_{eh+}           &J_{eh-}           &0           &0           \cr
   J_{eh-}           &J_{eh+}           &0           &0           \cr
   0           &0           &J_{eh+}           &J_{eh-}           \cr
   0           &0           &J_{eh-}           &J_{eh+}          
\end{pmatrix}
\nonumber \\
+\frac{\alpha _{eh}}{2}
&\begin{pmatrix}
   0            &0            &J_{eh+}           &J_{eh-}           \cr
   0            &0            &J_{eh-}           &J_{eh+}           \cr
   J_{eh+}           &J_{eh-}           &0             &0           \cr
   J_{eh-}           &J_{eh+}           &0             &0           
\end{pmatrix}  , 
\label{eq:H4}
\end{align}
where 
$J_{eh \pm}=J_{eh1} \pm J_{eh2}$
and 
\begin{align}
 J_{eh2}=&\left( t_A ^2 +t_B ^2 \right)
 \Bigl( 
    \frac{f_B^2}{U'-\Delta +\Delta_J-2J}+\frac{g_B ^2}{U'-\Delta -\Delta_J-2J}\Bigr) . 
\label{143402_2Aug10}
\end{align}
The basis set is 
$\{ \ket{\psi_{e \downarrow}, \psi_{h \uparrow}} , \ket{ \psi_{e \uparrow}, \psi_{h \downarrow}},$ $\ket{\psi_{h \uparrow}, \psi_{e \downarrow}}, \ket{\psi_{h \downarrow}, \psi_{e \uparrow}} \}$. 

(5) $(n=5,\ S_z=3/2)$
\begin{align}
{\cal H}^{(5,3/2)}=
\begin{pmatrix}
   0 &t_B \cr
   t_B & 0
\end{pmatrix} .
\label{201858_14Nov10}
\end{align}
The basis set is 
$\{ \ket{\psi_{e \uparrow}, \psi_{H+1}},  \ket{\psi_{H+1}, \psi_{e \uparrow}} \} $. 

(6) $(n=5,\ S_z=1/2)$
\begin{align}
{\cal H}^{(5,1/2)}=\ 
\begin{pmatrix}
 -J_{eH}          &\frac{J_{eH}}{\sqrt{2}} & 0                &-\frac{t_B}{\sqrt{2}}      & 0                & 0                 \cr
  J_{eH}\sqrt{2}  &-\frac{J_{eH}}{2}       &-\frac{t_B}{\sqrt{2}}     &-\frac{t_B}{2}     & 0                & 0                 \cr
 0                 &-\frac{t_B}{\sqrt{2}}    &-J_{eH}          & \frac{J_{eH}}{\sqrt{2}}  & 0                & 0                 \cr
 -\frac{t_B}{\sqrt{2}}     &-\frac{t_B}{2}           & J_{eH}\sqrt{2}  &-\frac{J_{eH}}{2} & 0                & 0                 \cr
 0                 & 0               & 0                & 0               &-J_{eL}& -f_B ^2 t_A      \cr
 0                 & 0               & 0                & 0               & -f_B ^2 t_A      &-J_{eL}
\end{pmatrix}.
\label{161632_2Aug10}
\end{align}
where 
\begin{align}
J_{eH}&=\frac{t_B ^2}{4J}+\frac{t_A^2}{U+U'+2J}  \nonumber\\
   &+\frac{t_A^2 f_B ^2}{\Delta+U-U'+J-\Delta_J} \nonumber\\
   &+\frac{t_A^2 g_B ^2}{\Delta+U-U'+J+\Delta_J} ,
\label{161730_2Aug10}
\end{align}
and 
\begin{align}
J_{eL}&=\frac{t_A^2 f_B  ^2g_B ^2}{2\Delta_J}+\frac{t_A^2 f_A ^2}{\Delta +2U'+J+\Delta_J} \nonumber\\
   & +\frac{f_A ^2 t_B  ^2}{2} \frac{3}{\Delta-U+U'+\Delta_J-J} \nonumber\\
   & +\frac{f_A ^2 t_B  ^2}{2} \frac{1}{\Delta -U+U'+\Delta_J+J} . 
\end{align}
The basis set is 
$\{ 
\ket{\psi_{e\downarrow}, \psi_{H+1}}, 
\ket{\psi_{e\uparrow}, \psi_{0}},
\ket{\psi_{H+1}, \psi_{e\downarrow}},$ 
$
\ket{\psi_{H0}, \psi_{e \uparrow}}, 
\ket{\psi_{e \uparrow}, \psi_{L}}, 
\ket{\psi_L, \psi_{e \uparrow}} 
\} $. 

\noindent
$^{\dagger}$Present address: 
Department of Physics, Hokkaido University, Sapporo 060-0810, Japan.


\begin{references}

\bibitem{Maekawa}
S.~Maekawa, T.~Tohyama, S.~E.~Barnes, S.~Ishihara, W.~Koshibae, and G.~Khliullin, 
{\it Physics of Transition Metal Oxides}
(Springer Verlag, Berlin, 2004). 

\bibitem{Nasu}
K.~Nasu,
{\it Photo Induced Phase Transition}
(World Scientific, Singapore, 2004), and references therein.

\bibitem{Fiebig}
M.~Fiebig, K.~Miyano, Y.~Tomioka, and Y.~Tokura,
Science {\bf 280}, 1925 (1998). 

\bibitem{Cavalleri}
A.~Cavalleri, Cs.~T\'{o}th, C.~W.~Siders, J.~A.~Squier,
F.~R\'{a}ksi, P.~Forget, and J.~C.~Kieffer,
Phys. Rev. Lett. {\bf 87}, 237401 (2001).

\bibitem{Okamoto}
H.~Okamoto, T.~Miyagoe, K.~Kobayashi, H.~Uemura, H.~Nishioka, H.~Matsuzaki, A.~Sawa, and Y.~Tokura,
Phys. Rev. B {\bf 83}, 125102 (2011).

\bibitem{Matsueda}
H.~Matsueda and S.~Ishihara, J. Phys. Soc. Jpn. {\bf 76}, 083703 (2007).

\bibitem{Kanamori1}
Y.~Kanamori, H.~Matsueda and S.~Ishihara,
Phys. Rev. Lett. {\bf 103}, 267401 (2009).
{\it ibid.},
Phys. Rev. B {\bf 82},
115101 (2010).

\bibitem{Iwai}
S.~Iwai, S.~Tanaka, K.~Fujinuma, H.~Kishida, H.~Okamoto, and Y.~Tokura,
Phys. Rev. Lett. {\bf 88}, 057402 (2002).

\bibitem{Chollet}
M.~Chollet, L.~Guerin, N.~Uchida, S.~Fukaya, H.~Shimoda, T.~Ishikawa, K.~Matsuda, T.~Hasegawa,
A.~Ota, H.~Yamochi, G.~Saito, R.~Tazaki, S.~Adachi, and S.~Koshihara,
Science {\bf 307}, 86 (2005). 

\bibitem{Tajima}
N.~Tajima, J.~Fujisawa, N.~Naka, T.~Ishihara, R.~Kato, Y.~Nishio, and K.~Kajita,
J. Phys. Soc. Jpn. {\bf 74}, 511 (2005).

\bibitem{Yonemitsu}
K.~Yonemitsu and K.~Nasu,
J. Phys. Soc. Jpn. {\bf 75}, 011008 (2006).

\bibitem{Sato1}
O.~Sato, T.~Iyoda, A.~Fukushima, and K.~Hashimoto,
Science {\bf 272} 704 (1996).

\bibitem{Bleuzen}
A.~Bleuzen, C.~Lomenech, V.~Escax, F.~Villain, F.~Varret, C.~C.~dit~Moulin, and M.~Verdaguer,
J. Am. Chem. Soc. {\bf 122} 6648 (2000).

\bibitem{Escax}
V.~Escax, A.~Bleuzen, C.~C.~dit~Moulin, F.~Villain, A.~Goujon, F.~Varret, and M.~Verdaguer,
J. Am. Chem. Soc. {\bf 123} 12536 (2001).

\bibitem{Sato2}
O.~Sato,
J. Photoch. Photobio. C: Photochem. Rev. {\bf 5} 203 (2004).

\bibitem{Willenbacher}
N.~Willenbacher and H.~Spiering,
J. Phys. C {\bf 21}, 1423 (1988).

\bibitem{Tchougreeff}
A.~L.~Tchougr\'{e}eff and M.~B.~Darkhovskii,
Int. J. Quantum Chem. {\bf 57}, 903 (1996).

\bibitem{Nishino}
M.~Nishino, K.~Boukheddaden, Y.~Konishi, and S.~Miyashita,
Phys. Rev. Lett. {\bf 98}, 247203 (2007).

\bibitem{Miyashita}
S.~Miyashita, P.~A.~Rikvold, T.~Mori, Y.~Konishi, M.~Nishino and H.~Tokoro,
Phys. Rev. B {\bf 80}, 064414 (2009).

\bibitem{Frontera}
C.~Frontera, J.~L.~Garc\'{i}a-Mu\~{n}oz, A.~Llobet, and M.~A.~G.~Aranda,
Phys. Rev. B {\bf 65}, 180405(R) (2002). 

\bibitem{Tsubouchi}
S.~Tsubouchi, T.~Kyomen, M.~Itoh, P.~Ganguly, M.~Oguni, Y.~Shimojo, Y.~Morii, and Y.~Ishii,
Phys. Rev. B {\bf 66}, 052418 (2002). 

\bibitem{Okimoto1}
Y.~Okimoto, X.~Peng, M.~Tamura, T.~Morita, K.~Onda, T.~Ishikawa, S.~Koshihara, N.~Todoroki, T.~Kyomen, and M.~Itoh,
Phys. Rev. Lett. {\bf 103}, 027402 (2009).

\bibitem{Asai}
K.~Asai, P.~Gehring, H.~Chou, and G.~Shirane,
Phys. Rev. B {\bf 40}, 10982 (1989).

\bibitem{Tokura}
Y.~Tokura, Y.~Okimoto, S.~Yamaguchi, H.~Taniguchi, T.~Kimura, and H.~Takagi,
Phys. Rev. B {\bf 58}, 1699(R) (1998).

\bibitem{Heikes}
R.~R.~Heikes, R.~C.~Miller, and R.~Mazelsky,
Physica {\bf 30}, 1600 (1964).

\bibitem{Raccah}
P.~M.~Raccah, and J.~B.~Goodenough,
Phys. Rev. {\bf 155}, 932 (1967).

\bibitem{Yamaguchi}
S.~Yamaguchi, Y.~Okimoto, and Y.~Tokura,
Phys. Rev. B {\bf 54}, 11022(R) (1996).

\bibitem{Saitoh}
T.~Saitoh, T.~Mizokawa, A.~Fujimori, M.~Abbate, Y.~Takeda, and M.~Takano,
Phys. Rev. B {\bf 55}, 4257 (1997).

\bibitem{Rao}
C.~N.~R.~Rao, Om~Parkash, D.~Bahadur, P.~Ganguly, and S.~Nagabhushana,
J. Solid State Chem. {\bf 22}, 353 (1977).

\bibitem{Rodrigouez}
M.~A.~Se\~{n}ar\'{i}s-Rodr\'{i}guez and J.~B.~Goodenough,
J. Solid State Chem. {\bf 118}, 323 (1995).

\bibitem{Itoh}
M.~Itoh and I.~Natori,
J. Phys. Soc. Jpn. {\bf 64}, 970 (1995).

\bibitem{Tutsui}
K.~Tsutsui, J.~Inoue, and S.~Maekawa,
Phys. Rev. B {\bf 59}, 4549 (1999).

\bibitem{Suzuki}
R.~Suzuki, T.~Watanabe, and S.~Ishihara,
Phys. Rev. B {\bf 80}, 054410 (2009).

\bibitem{Okimoto2}
Y.~Okimoto, T.~Miyata, M.~S.~Endo, M.~Kurashima, K.~Onda, T.~Ishikawa, S.~Koshihara,
M.~Lorenc, E.~Collet, H.~Cailleau, and T.~Arima, 
Phys. Rev. B {\bf 84}, 121102(R) (2011).

\bibitem{Iwai2}
S.~Iwai, S.~Tomimoto, Y.~Okimoto, J.~P.~He, Y.~Kaneko, H.~Okamoto, and Y.~Tokura,
Meeting abstracts of the Physical Society of Japan, {\bf 59}(2-4), 673.

\bibitem{Kanamori2}
Y.~Kanamori, H.~Matsueda, and S.~Ishihara,
Phys. Rev. Lett. {\bf 107}, 167403 (2011).

\bibitem{Khomskii}
D.~I.~Khomskii and U.~L\"{o}w,
Phys. Rev. B {\bf 69}, 184401 (2004).

\bibitem{Mclachlan}
A.~D.~McLachlan and M.~A.~Ball,
Rev. Mod. Phys. {\bf 36}, 844 (1964).

\bibitem{Terai}
A.~Terai and Y.~Ono,
Prog. Theor. Phys. Suppl. {\bf 113}, 177 (1993).

\end{references}
\end{document}